\documentclass[a4paper,fleqn,usenatbib]{mnras}

\usepackage{mathptmx}

\usepackage[T1]{fontenc}
\usepackage{ae,aecompl}

\usepackage{graphicx}	
\usepackage{amsmath}	
\usepackage{amssymb}	

\title[Dust and molecules in galaxies]{Effects of dust evolution on the abundances
of CO and H$_2$}

\author[H. Hirashita and N. Harada]{
Hiroyuki Hirashita$^{1}$\thanks{E-mail: hirashita@asiaa.sinica.edu.tw}
and Nanase Harada$^{1}$
\\
$^{1}$Institute of Astronomy and Astrophysics, Academia Sinica,
PO Box 23-141, Taipei 10617, Taiwan
}

\date{Accepted XXX. Received YYY; in original form ZZZ}

\pubyear{2016}

\begin{document}
\label{firstpage}
\pagerange{\pageref{firstpage}--\pageref{lastpage}}
\maketitle

\begin{abstract}
The CO-to-H$_2$ conversion factor ($X_\mathrm{CO}$) is known to
correlate with the metallicity ($Z$). The dust abundance, which is
related to the metallicity, is responsible for this correlation through
dust shielding of dissociating photons and H$_2$ formation on dust surfaces.
In this paper, we investigate
how the relation between dust-to-gas ratio and metallicity
($\mathcal{D}$--$Z$ relation)
affects the H$_2$ and CO abundances (and $X_\mathrm{CO}$)
of a `molecular' cloud.
For the $\mathcal{D}$--$Z$ relation, we adopt a dust evolution
model developed in our previous
work, which treats the evolution of not only dust abundance but
also grain sizes in a galaxy.
Shielding of dissociating photons and H$_2$ formation on dust
are solved consistently with the dust abundance and
grain sizes. As a consequence, our models {predict consistent metallicity dependence of
$X_\mathrm{CO}$ with observational data}. Among various processes driving dust evolution,
grain growth by accretion has the largest
impact on the $X_\mathrm{CO}$--$Z$ relation.
The other processes also have some impacts on the $X_\mathrm{CO}$--$Z$ relation,
but their effects are minor compared with the scatter of the observational data
at the metallicity range ($Z\ga 0.1$~Z$_{\sun}$) where CO could be detected.
We also find that dust condensation in stellar ejecta has a dramatic
impact on the H$_2$ abundance at low metallicities
($\la 0.1$ Z$_{\sun}$), relevant for damped Lyman $\alpha$ systems and
nearby dwarf galaxies, and that the grain size dependence of H$_2$ formation
rate is also important.
\end{abstract}




\section{Introduction}

Galaxies evolve through star formation. Since
molecular clouds are the birth place of stars,
understanding how molecular clouds evolve provides us with
an important key to how stars form in galaxies. The main chemical
constituent of molecular clouds is molecular
hydrogen (H$_2$). Since emission from H$_2$ is
weak in cold molecular environments
(H$_2$ emission is more easily observed in regions with
shock or ultraviolet excitation; e.g.\ \citealt{Naslim:2015aa}
for a recent observation),
emission from carbon monoxide (CO) is often used as a tracer of molecular clouds.
Thus, understanding the formation and evolution of
H$_2$ and CO and the relation between
these two species is important to clarify to what extent
CO can really trace molecular clouds at various stages of
galaxy evolution.

In the present Universe, H$_2$ forms predominantly on the surface of
dust grains \citep{Gould:1963aa}.
On the other hand, CO forms in the gas phase via various reactions.
Both species favour `shielded' environments with a high
column density of gas since
they are dissociated by ultraviolet (UV) radiation.
At high column densities, H$_2$ molecules can shield
UV radiation by their own absorption \citep{Draine:1996aa}
(i.e.\ self-shielding). Since self-shielding of CO is weaker
\citep{Lee:1996aa}, CO forms in more
embedded regions than H$_2$. Dust extinction also
plays an important role in shielding dissociating radiation.
Although CO emission is empirically known to trace H$_2$ in
solar-metallicity environments,
it is not obvious that this is generally true for galaxies
(i) because these species have
different formation mechanisms and the formation rates have
different dependence on
metallicity \citep{Maloney:1988aa}, and
(ii) because UV intensity and dust abundance vary in different
stages of galaxy evolution.

For the purpose of deriving the H$_2$ abundance from an observed CO
intensity, one assumes a CO-to-H$_2$ conversion factor,
$X_\mathrm{CO}\equiv N_\mathrm{H_2}/W_\mathrm{CO}$,
where $N_\mathrm{H_2}$ is the H$_2$ column density,
and $W_\mathrm{CO}$ is the intensity of the CO
$J=1\to 0$ emission line integrated for the frequency
(the frequency is often converted to the velocity shift
in units of km~s$^{-1}$). Another expression for the conversion
factor, $\alpha_\mathrm{CO}$, is based on
{the column mass density (or surface mass density) of the molecular
gas, $\Sigma_\mathrm{mol}=1.36m_\mathrm{H}(2N_\mathrm{H_2})$,
where 1.36 is a factor to account for the contribution of helium to the
total mass. In this expression, the conversion factor is defined as
$\alpha_\mathrm{CO}\equiv\Sigma_\mathrm{mol}/W_\mathrm{CO}$.}
In this paper, we represent the CO-to-H$_2$ conversion factor
by $X_\mathrm{CO}$.

Deriving a reasonable CO-to-H$_2$ conversion factor is
not easy. To observationally derive
the CO-to-H$_2$ conversion
factor, we need to know the H$_2$ content, which is not
directly observed by its emission. The H$_2$ content
(or the total column density or mass of a molecular cloud) is
usually estimated indirectly through
an estimate of the virial mass or a conversion from the dust far-infrared
intensity to the total gas column density
\citep[see][for a review]{Bolatto:2013aa}.
With such methods of obtaining the H$_2$ content,
$X_\mathrm{CO}$ has been derived for various
galaxies, mainly nearby ones.
In particular, it has been found that $X_\mathrm{CO}$
depends on the metallicity
\citep{Wilson:1995aa,Arimoto:1996aa,Israel:1997aa,Bolatto:2008aa,Leroy:2011aa,Hunt:2015aa}
as well as the density and temperature
(\citealt{Feldmann:2012aa}, hereafter FGK12; \citealt{Narayanan:2012aa}).

As mentioned above, dust has an influence on both
H$_2$ and CO abundances through shielding of UV dissociating
photons and H$_2$ formation on dust surfaces.
The metallicity dependence of $X_\mathrm{CO}$ mentioned
above may reflect dust enrichment, which occurs simultaneously
with metal enrichment \citep{Lisenfeld:1998aa,Dwek:1998aa}.
Dust enrichment is one of the most important aspects
for understanding the evolution of galaxies.
Dust absorbs and scatters the stellar light and emit it
in the far infrared, thereby
shaping the spectral energy distribution
\citep[e.g.][for recent modelling]{Yajima:2014aa,Schaerer:2015aa}.
As mentioned above, grain surfaces provide a condition suitable for 
efficient formation of molecular hydrogen, which is an important 
coolant in low-metallicity clouds \citep[e.g.][]{Cazaux:2004aa}. 
Dust itself is also an important coolant in star formation,
inducing the final fragmentation that determines the stellar mass 
\citep{Omukai:2005aa,Schneider:2006aa}.

Not only the dust abundance, but also
the grain size distribution is important in various aspects.
In particular,
shielding of dissociating photons and H$_2$ formation on dust
surfaces depend on the grain size distribution through the
dependence of the surface area on the grain size.
\citet{Asano:2013aa} formulated the evolution of
grain size distribution in a consistent manner with galaxy evolution.
In their calculation, dust condensed in stellar ejecta
[supernovae (SNe) and asymptotic giant branch star winds],
dominate the grain size distribution at the early stage of galactic
evolution \citep[see also][]{Valiante:2009aa}.
These stellar sources form large ($\sim 0.1~\micron$)
grains, based on theoretical and observational evidence
(see section 2.1 of \citealt[][hereafter H15]{Hirashita:2015ab}, for detailed references); thus, the
dust is dominated by large grains at the early stage of galaxy evolution.
As the system is enriched with dust,
shattering as a result of grain--grain collision becomes efficient
enough to increase the abundance of small grains.
The increase of small grains drastically boosts the
total grain surface area; as a consequence, grain growth by
the accretion of gas-phase metals becomes the most 
important process for dust enrichment. Afterwards, the
abundant small grains coagulate to form large grains.
\citet{Nozawa:2015aa} used the same model with a modification
of incorporating the molecular cloud phase, which hosts
strong accretion and coagulation, in addition to the originally
included warm and cold phases. They explained not only
the Milky Way extinction curve but also the extinction curve observed
in a high-redshift quasar taken by \citet{Maiolino:2004aa}
\citep[see also][]{Gallerani:2010aa}.

Although the above recent models took into account the full details
of grain formation and processing mechanisms,
there are still some uncertain free parameters regarding, especially,
the time-scales
of individual processes. The time-scales (or efficiencies) of
accretion, shattering, and coagulation are strongly affected by
the density structures in the ISM \citep{Bekki:2015ab,McKinnon:2016aa,Aoyama:2016aa}, since
accretion and coagulation take place only in the dense and cold ISM
while shattering occurs predominantly in the diffuse ISM
\citep{Yan:2004aa,Asano:2013aa}. Therefore, to complement those detailed
models, a parameter survey study is desirable; that is,
we need to survey all the reasonable ranges
of the time-scales, for the purpose of examining how
sensitive the evolution of grain size distribution is
to the assumed time-scales
or for the purpose of finding the ranges of the time-scales
that reproduce successfully the observed extinction curves
\citep{Bekki:2015aa,Hou:2016aa}.
However,  a full treatment of grain
size distribution requires a lot of computational time,
and is not suitable for such a parameter survey study.

To make a parameter survey possible in a reasonable computational
time, we adopt a simplified model developed by H15
to calculate the
evolution of grain size distribution: H15 adopted a `two-size
approximation' approach, in which the grain sizes are
represented by two sizes (large and small grains) separated
around a radius of 0.03~$\micron$.
This approximated model still includes all the above processes considered by
\citet{Asano:2013aa} but simply treats the production of or
mass exchange between
the small and large grain populations for the evolution of grain size
distribution. H15 showed that
this two-size approximation traces the same evolutionary
behaviours of grain size distribution and extinction curve
as presented in
\citet{Asano:2013aa,Asano:2014aa}. Therefore,
H15 confirmed that the two-size approximation can be used
as a simplified (or computationally cheap) version of
the full treatment of grain size distribution. \citet{Bekki:2015aa}
applied the two-size approximation in combination with
extinction curve calculations, finding that the variety of
extinction curves among the Milky Way and the Large and
Small Magellanic Clouds can be explained by different
transportation efficiencies of small grains out of the galaxies.
Indeed, they took advantage of the quickness of two-size
approximation calculations to find optimum parameters
that reproduce the observed extinction curves
\citep[see also][]{Hou:2016aa}.

In this paper, utilizing the dust models mentioned above
(specifically the two-size approximation
in H15), we examine the effect of dust evolution on
the abundances of H$_2$ and CO molecules.
This enables us to examine the effect of dust evolution on
the H$_2$ and CO abundances. In particular, we
will focus on how the dust evolution, including the evolution of
grain sizes, affects the metallicity dependence of $X_\mathrm{CO}$.
The dependence of H$_2$ formation rate on the evolution of
grain size distribution has not been fully investigated:
\citet{Yamasawa:2011aa} incorporated the
effect of grain size distribution on H$_2$ formation in a galaxy
evolution model, but they only focused on the early stage of
galaxy evolution. Since there have already been
analytic models for H$_2$ formation
\citep{Hirashita:2005aa,Krumholz:2008aa,Krumholz:2009aa,McKee:2010aa},
we focus on the effect of grain size, which has not been included in
those models.
The grain size distribution also affects the
CO abundance through shielding of dissociating photons.
Through this work, we not only understand or quantify the
effect of dust evolution on $X_\mathrm{CO}$ but also estimate
how much the observational scatter in the $X_\mathrm{CO}$--metallicity
relation can be explained by the variation of dust evolution among
galaxies. Since there are several processes driving the dust evolution,
using a method suitable for a parameter survey such as
the two-size approximation is crucial for the present work.

This paper is organized as follows.
In Section \ref{sec:model}, we model the dust evolution, the abundances
of H$_2$ and CO, and the CO-to-H$_2$ conversion factor.
In Section \ref{sec:result}, we show the results for the basic predictions of
our models. In Section \ref{sec:discussion}, we further discuss the
dependence on some physical components that cause major
influences on the H$_2$ and CO abundances. In Section
\ref{sec:implication}, we comment on some implications for various
galaxy populations. Finally we give conclusions in Section \ref{sec:conclusion}.
Throughout this paper, we adopt the typical conversion factor for the
Milky Way as
$X_\mathrm{CO}=2\times 10^{20}$ cm$^{-2}$ K$^{-1}$ km$^{-1}$ s,
which corresponds to
$\alpha_\mathrm{CO}=4.3$ M$_{\sun}$ K$^{-1}$ km$^{-1}$ s pc$^2$
\citep{Bolatto:2013aa},
and $Z_{\sun}$ = 0.02 for the solar metallicity used for the metallicity
normalization in the models (we adopt the same value as in H15).

\section{Model}\label{sec:model}

We first calculate how the dust content evolves as a function of
metallicity to obtain the relation between dust-to-gas ratio and
metallicity ($\mathcal{D}$--$Z$ relation).
In this calculation, we also trace the evolution of grain sizes
by adopting the two-size approximation described below.
Using the computed $\mathcal{D}$--$Z$ relation,
we calculate the molecular abundance of a single typical
cloud in the galaxy. As a consequence, the
model gives how the molecular abundances in a cloud is affected
by the $\mathcal{D}$--$Z$ relation.

\subsection{Dust evolution}\label{subsec:dust}

We use the two-size dust enrichment model developed by H15.
In this model, to avoid heavy computation,
the entire size range of dust grains is represented by small and
large grains (roughly divided at $a\sim 0.03~\micron$, where $a$
is the grain radius), considering
that various grain processing mechanisms work differently between these two
grain populations. The model takes into account dust condensation in
stellar ejecta, dust destruction in SN shocks,
grain growth by accretion and coagulation, and grain disruption by shattering. 

H15 considered the evolution of the abundances of small and large grains,
and calculated
the total mass of small/large grains divided by the total gas mass,
referred to as the small/large-grain dust-to-gas ratio.
We denote the small- and large-grain dust-to-gas ratios as
$\mathcal{D}_\mathrm{s}$ and $\mathcal{D}_\mathrm{l}$, respectively.
The total dust-to-gas ratio is the sum of these two components as
$\mathcal{D}=\mathcal{D}_\mathrm{s}+\mathcal{D}_\mathrm{l}$.
After applying the instantaneous recycling approximation \citep{Tinsley:1980aa},
the evolutions of $\mathcal{D}_\mathrm{s}$ and $\mathcal{D}_\mathrm{l}$ are
described in terms of metallicity evolution as (see H15 for the derivation)
\begin{align}
\mathcal{Y}_Z\frac{\mathrm{d}\mathcal{D}_\mathrm{l}}{\mathrm{d}Z} &=
f_\mathrm{in}(\mathcal{R}Z+\mathcal{Y}_Z)+\beta_\mathrm{co}
\mathcal{D}_\mathrm{s}-(\beta_\mathrm{SN}+\beta_\mathrm{sh}
+\mathcal{R})\mathcal{D}_\mathrm{l},\label{eq:large}\\
\mathcal{Y}_Z\frac{\mathrm{d}\mathcal{D}_\mathrm{s}}{\mathrm{d}Z} &=
\beta_\mathrm{sh}\mathcal{D}_\mathrm{l}-\left(
\beta_\mathrm{SN}+\beta_\mathrm{co}+\mathcal{R}-
\beta_\mathrm{acc}\right)\mathcal{D}_\mathrm{s},\label{eq:small}
\end{align}
where $Z$ is the metallicity,
$f_\mathrm{in}$ is the dust condensation efficiency of metals in the stellar ejecta,
{$\mathcal{R}$ is the returned fraction of gas from stars,}
$\mathcal{Y}_Z$ is the mass fraction of newly produced metals by stars,
and $\beta_\mathrm{SN}$, $\beta_\mathrm{sh}$, $\beta_\mathrm{co}$ and 
$\beta_\mathrm{acc}$ indicate the efficiencies (explained below) of 
SN shock destruction, shattering, coagulation and accretion, respectively.
Note that these efficiencies (referred to as $\beta$s) depend on
$\mathcal{D}_\mathrm{s}$ or $\mathcal{D}_\mathrm{l}$ except
$\beta_\mathrm{SN}$ (H15).
The right-hand size of equation (\ref{eq:large}) represents
the stellar dust production
[$f_\mathrm{in}(\mathcal{R}Z+\mathcal{Y}_Z)$; note that dust grains supplied by
stars are assumed to be large; see H15 and references therein],
the increase by coagulation
of small grains ($\beta_\mathrm{co}\mathcal{D}_\mathrm{s}$),
the decreases by SN destruction and shattering
($\beta_\mathrm{SN}\mathcal{D}_\mathrm{l}$ and
$\beta_\mathrm{sh}\mathcal{D}_\mathrm{l}$, respectively),
and the dilution of dust-to-gas ratio by returned gas
from stars ($\mathcal{RD}_\mathrm{l}$). Equation 
(\ref{eq:small}) shows the increase of small grains:
the shattering and coagulation terms in
equation (\ref{eq:small}) have
the opposite sign to those in equation (\ref{eq:large}),
since shattering is source and coagulation is sink for
small grains. The increase of dust abundance by accretion
($\beta_\mathrm{acc}\mathcal{D}_\mathrm{s}$) only appears
in equation (\ref{eq:small}) based on the argument that
accretion is much more efficient for small grains than 
for large grains because small grains have much larger
surface-to-volume ratio \citep{Hirashita:2011aa,Asano:2013aa}.

The efficiencies denoted as $\beta$s are defined as
$\beta_\mathrm{SN}\equiv\tau_\mathrm{SF}/\tau_\mathrm{SN}$,
$\beta_\mathrm{sh} \equiv \tau_\mathrm{SF}/ \tau_\mathrm{sh}$ and $\beta_\mathrm{co} \equiv \tau_\mathrm{SF}/ \tau_\mathrm{co}$,
where $\tau_\mathrm{SF}\equiv M_\mathrm{gas}/\psi$
($M_\mathrm{gas}$ is the total gas mass in the galaxy and $\psi$ is
the star formation rate) is the star formation time-scale,
and the shattering and coagulation time-scales are proportional to
the dust-to-gas ratios, since shattering and coagulation are collisional
processes:
\begin{align}
\tau_\mathrm{sh} = \tau_\mathrm{sh,0}\left(
\frac{\mathcal{D}_\mathrm{l}}{\mathcal{D}_\mathrm{MW,l}}\right)^{-1},
\label{eq:sha}\\
\tau_\mathrm{co} = \tau_\mathrm{co,0}\left(
\frac{\mathcal{D}_\mathrm{s}}{\mathcal{D}_\mathrm{MW,s}}\right)^{-1}, 
\label{eq:co}
\end{align}
where the time-scales are normalized to $\tau_\mathrm{sh,0}$ and 
$\tau_\mathrm{co,0}$, which are the values at the MW dust-to-gas ratio, 
$\mathcal{D}_\mathrm{MW,l} = 0.007$  and $\mathcal{D}_\mathrm{MW,s} = 0.003$ 
(H15). The accretion efficiency
$\beta_\mathrm{acc}$ is regulated by the lifetime of dense
clouds ($\tau_\mathrm{cl}$) and the metallicity. Accretion is more efficient for
longer $\tau_\mathrm{cl}$. For $\beta_\mathrm{acc}$,
we need the averaged values of $a^\ell$ ($\ell =1$, 2, and 3)
 (see Appendix \ref{app:moment})
because accretion is a surface process in which the grain size,
surface and volume play an important role. The calculation of
$\beta_\mathrm{acc}$ is explained in Appendix \ref{app:beta_acc}.

By solving equations (\ref{eq:large}) and (\ref{eq:small}), we obtain
the relation between dust-to-gas ratio and metallicity
($\mathcal{D}$--$Z$ relation).
Below, we consider a cloud whose column density is typical of
Galactic molecular clouds, and calculate the H$_2$ and CO
abundances in the cloud based on the dust abundance given
as a function of metallicity by the above $\mathcal{D}$--$Z$ relation.

\subsection{H$_2$ abundance}\label{subsec:H2}

We consider a cloud with a hydrogen column density of
$N_\mathrm{H}\sim 10^{22}$~cm$^{-2}$ (typical of Galactic
molecular clouds) under a given metallicity.
We assume that this cloud has small- and large-grain
dust-to-gas ratios $\mathcal{D}_\mathrm{s}(Z)$ and
$\mathcal{D}_\mathrm{l}(Z)$ calculated by the model in
Section~\ref{subsec:dust}.
The H$_2$ abundance is assumed to be determined by
the balance between formation and dissociation
(this equilibrium assumption is discussed at the end of this
subsection).
The H$_2$ formation rate is proportional to
the local density represented by the
number density of hydrogen nuclei $n_\mathrm{H}$.
For dissociation, the column density of
hydrogen nuclei $N_\mathrm{H}$ as well as the UV radiation
field $\chi$
is important for shielding of dissociating photons
[$\chi$ is the UV radiation field intensity at the Lyman-Werner (LW)
band normalized to the solar neighbourhood value
derived by \citealt{Habing:1968aa},
$3.2\times 10^{-20}$ erg s$^{-1}$ cm$^{-2}$ Hz$^{-1}$ sr$^{-1}$;
see also \citealt{Hirashita:2005aa}].
{Expressing the UV radiation field with a single parameter
$\chi$ means that we neglect the dependence of stellar UV spectrum
(or hardness)
on metallicity. As shown later, the variation of UV radiation has a large
influence on the molecular abundances at low metallicity, where
we are not much interested in the CO abundance since it is too low
to be detected. The H$_2$ abundance at
low metallicity could be affected by the variation of UV spectrum.
However, according to \citet{Schaerer:2002aa}, with a fixed star formation rate,
the H$_2$ dissociation rate only differs by a factor of 1.5 between
low ($\sim 1/50$ Z$_{\sun}$) and solar metallicities, while
we will vary $\chi$ by orders of magnitude in this paper. Thus, we
simply concentrate on the variation of $\chi$ (UV intensity) and neglect the
variation of hardness.}

Under a given set of $(n_\mathrm{H},\, N_\mathrm{H},\, \chi )$,
we calculate the H$_2$ fraction, $f_\mathrm{H_2}$ achieved
as a result of the equilibrium between the formation on dust
and the dissociation by UV (LW band) radiation.
{The H$_2$ fraction is defined in such a way that
$f_\mathrm{H_2}N_\mathrm{H}/2$ gives the column density of
H$_2$.}
The rates of formation and dissociation are given in what follows.

The increasing rate of $f_\mathrm{H_2}$ by H$_2$ formation on dust
is described as
\begin{align}
\left[\frac{\mathrm{d}f_\mathrm{H_2}}{\mathrm{d}t}\right]_\mathrm{form}=2
\sum_i(1-f_\mathrm{H_2})R_\mathrm{H_2,dust}^in_\mathrm{H},
\end{align}
where
the rate coefficient $R_\mathrm{H_2,dust}^i$ (superscript $i$
indicates small or large grains; i.e.\ $i=\mathrm{s}$ or
$i=\mathrm{l}$) is introduced.
The rate coefficient, which depends on the dust-to-gas ratio,
is determined by \citep{Yamasawa:2011aa}
\begin{align}
R_\mathrm{H_2,dust}^i=
\frac{3\mathcal{D}_i\mu m_\mathrm{H}S_\mathrm{H}\bar{v}\langle a^2\rangle_i}{8s\langle a^3\rangle_i},
\label{eq:H2_form}
\end{align}
where $s$ is the material density of dust,
$m_\mathrm{H}$ is the atomic
mass of hydrogen,
$\bar{v}$ is the mean thermal speed, $S_\mathrm{H}$ is the
probability of a hydrogen atom reacting with
another hydrogen atom on the dust surface to form H$_2$, and
$\langle a^2\rangle_i$ and $\langle a^3\rangle_i$ are the second and
third moments with subscript $i$ indicating small or large grains
(Appendix \ref{app:moment}).
We adopt
$s=3.3$ g cm$^{-3}$ \citep{Draine:1984aa} and $\mu =1.4$.
Adopting instead $s\sim 2$ g cm$^{-3}$ appropriate for carbonaceous
dust does not affect our conclusions below.
We also fix $S_\mathrm{H}=0.3$: such a high value is reasonable to
adopt since we are particularly interested in H$_2$ formation
in cold and shielded environments \citep{Hollenbach:1979aa}.
The thermal speed is estimated
as \citep{Spitzer:1978aa}
\begin{align}
\bar{v}=\sqrt{\frac{8k_\mathrm{B}T_\mathrm{gas}}{\pi m_\mathrm{H}}},
\end{align}
where $k_\mathrm{B}$ is the Boltzmann constant, and
$T_\mathrm{gas}$ is the kinetic temperature of the gas.

The changing rate of $f_\mathrm{H_2}$ by photodissociation is estimated as
\begin{align}
\left[\frac{\mathrm{d}f_\mathrm{H_2}}{\mathrm{d}t}\right]_\mathrm{diss}
=-R_\mathrm{diss}f_\mathrm{H_2},
\end{align}
where the rate coefficient $R_\mathrm{diss}$ is given by
\citep{Hirashita:2005aa}
\begin{align}
R_\mathrm{diss}=4.4\times 10^{-11}\chi S_\mathrm{shield,H_2}
S_\mathrm{shield,dust},
\end{align}
with $S_\mathrm{shield,H_2}$
and $S_\mathrm{shield,dust}$ being the correction factors for H$_2$ self-shielding and dust
extinction, respectively. We adopt the following form for $S_\mathrm{shield}$
\citep{Draine:1996aa,Hirashita:2005aa}:
\begin{align}
S_\mathrm{shield,H_2}=\min\left[1,\,\left(
\frac{\frac{1}{2}f_\mathrm{H_2}N_\mathrm{H}}{10^{14}~\mathrm{cm}^{-2}}
\right)^{-0.75}\right] ,
\end{align}
and
\begin{align}
S_\mathrm{shield,dust}=\exp\left(-\sum_i\tau_{\mathrm{LW},i}\right) ,
\label{eq:shield_dust}
\end{align}
where $\tau_{\mathrm{LW},i}$ is the optical depth of
dust component $i$ at the LW band. This optical depth is
estimated as
\begin{align}
\tau_{\mathrm{LW},i}=
\left(\frac{\tau_\mathrm{LW}}{N_\mathrm{H}}\right)_{0.01,i}
\left(\frac{\mathcal{D}_i}{0.01}\right) N_\mathrm{H} ,\label{eq:tau_LW}
\end{align}
where $({\tau_\mathrm{LW}}/{N_\mathrm{H}})_{0.01,i}$ is the
optical depth of dust component $i$ (again, $i=\mathrm{s}$ or l
for small and large grains, respectively)
for $\mathcal{D}_i=0.01$ at the LW band,
normalized to the
hydrogen column density.
H15 showed that, with
$\mathcal{D}_\mathrm{s}=0.003$ and $\mathcal{D}_\mathrm{l}=0.007$,
which are appropriate for the Milky Way, the Milky Way extinction
curve is reproduced with
a mass ratio of silicate-to-carbonaceous dust of 0.54 : 0.46.
We fix this silicate-to-carbonaceous dust ratio in calculating the
extinction for simplicity.
If we adopt the representative
wavelength for the LW band as 1000 \AA, we obtain
$({\tau_\mathrm{LW}}/{N_\mathrm{H}})_\mathrm{0.01,s}=8.2\times 10^{-21}$ cm$^2$
and
$({\tau_\mathrm{LW}}/{N_\mathrm{H}})_\mathrm{0.01,l}=2.3\times 10^{-21}$ cm$^2$.
{Fixing the silicate-to-carbonaceous dust abundance ratio does not
affect our results significantly. Grain composition is indeed suggested to be different in
different galaxies: studies on extinction curves indicate that
the dust composition in the Small Magellanic Cloud (SMC) is dominated by
silicate with little contribution from carbonaceous dust
\citep{Pei:1992aa,Weingartner:2001aa}.
Under the same grain size distribution, the extinction curve composed purely
of silicate
has an enhancement by a factor of $\sim 1.4$ around 1000 \AA\ compared
with the silicate-graphite mixture that reproduces the Milky Way extinction
curve \citep{Pei:1992aa}.
As we see later, dust extinction is important for CO formation,
but considering the sensitive dependence of CO abundance on
dust-to-gas ratio, difference in $\tau_\mathrm{LW}$ only by a factor
of 1.4 does not affect our discussions and conclusions in this paper.
Although the above extinction curve studies cannot exclude
a possibility of other dust species than silicate and carbonaceous dust
(note that \citealt{Tchernyshyov:2015aa} suggest that dust containing no
silicon atom should exist in some regions in the SMC based on
their study of elemental depletion),
we concentrate on the change of dust abundance $\mathcal{D}_i$
by fixing the coefficient $({\tau_\mathrm{LW}}/{N_\mathrm{H}})_{0.01,i}$
in this paper.}

In the above, we assumed the equilibrium between H$_2$ formation
and destruction. Thus, we implicitly assume
that the cloud has a longer lifetime than the H$_2$ formation time-scale.
The H$_2$ formation time-scale,
$\tau_\mathrm{H_2}$, is estimated as
\begin{align}
\tau_\mathrm{H_2}=\frac{f_\mathrm{H_2}}
{[\mathrm{d}f_\mathrm{H_2}/\mathrm{d}t]_\mathrm{form}}.
\end{align}
If we approximate the reaction rate in equation (\ref{eq:H2_form})
by using an intermediate value of
$\langle a^3\rangle_i/\langle a^2\rangle_i\sim 0.01~\micron$
between large and small grains (Table~\ref{tab:moments}) and
by replacing $\mathcal{D}_i$ with $\mathcal{D}$,
we obtain (omitting superscript $i$)
$R_\mathrm{H_2,dust}\sim 3.7\times 10^{-17}(\mathcal{D}/0.01)(T/10~\mathrm{K})^{1/2}
(n_\mathrm{H}/10^3~\mathrm{cm}^{-3})$
cm$^3$ s$^{-1}$, which leads to
$\tau_\mathrm{H_2}\sim f_\mathrm{H_2}/(R_\mathrm{H_2,dust}n_\mathrm{H})
\sim 8.7\times 10^5f_\mathrm{H_2}(\mathcal{D}/0.01)^{-1}(T/10~\mathrm{K})^{-1/2}
(n_\mathrm{H}/10^3~\mathrm{cm}^{-3})^{-1}$ yr.
This time-scale can be compared with the free-fall time estimated as
$\tau_\mathrm{ff}\sim 1.4\times 10^6(n_\mathrm{H}/10^3~\mathrm{cm}^{-3})^{-1/2}$ yr.
We find that $\tau_\mathrm{H_2}<\tau_\mathrm{ff}$ is satisfied if the dust-to-gas
ratio is near to the Milky Way value. Thus, if we consider a typical molecular
cloud, the equilibrium assumption on the H$_2$ abundance is reasonable,
since the time-scale of gravitational evolution occurs on a longer time-scale than
the reaction time-scale (although the reaction may be non-equilibrium for
small-scale structures; \citealt{Gibson:2015aa}).
However, if the dust-to-gas ratio is significantly lower than
the Milky Way value, the equilibrium assumption should be considered carefully,
since $\tau_\mathrm{H_2}$ becomes long in proportion to $\mathcal{D}^{-1}$
\citep[see also][]{Hu:2016aa}.
The assumption of equilibrium is yet justified if the molecular cloud is sustained
against the free fall or if the resulting $f_\mathrm{H_2}$ is significantly smaller
than unity.
{For example, if the molecular cloud is sustained for a few free-fall times
as indicated for nearby molecular clouds (e.g.\
\citealt{Ward-Thompson:2007aa}, although this is still debated), the equilibrium assumption
for H$_2$ formation is reasonable down to $\mathcal{D}\sim 0.001$.
In terms of the comparison with observations, we are mainly interested in
objects with $\mathcal{D}\ga 0.001$, for which CO could be detected.
Nevertheless, we should keep in mind that our results around
$\mathcal{D}\sim 0.001$, roughly corresponding to $Z\sim 0.1$ Z$_{\sun}$
are interpreted carefully, and that calculations taking into account
non-equilibrium H$_2$ formation is desirable for a future study.}

\subsection{CO abundance}\label{subsec:xco}

The CO abundance is determined by a complicated chemical network
that includes a lot of reactions. For example, \citet{Glover:2010aa}
treated a chemical network composed of $\sim 200$ reactions.
Since calculations of such a large chemical network are generally
time-consuming, performing full chemical calculations for the
CO abundance with a large number of cases for
dust evolution at various metallicities is not realistic.
Therefore, we utilize CO abundance data already calculated for
various physical conditions by \citet[][hereafter GM11]{Glover:2011aa}
(see also \citealt{Shetty:2011aa}). They
calculated spatially resolved H$_2$ and CO abundances
over a region of 5--20 pc
using dynamical simulations of magnetized turbulence coupled with
a chemical network for H$_2$ and CO and a treatment of
photodissociation. We follow the method used by FGK12, who
adopted the calculation results in GM11. Below we explain the
calculation method of CO abundance.

{Before we explain our formulation, we need to relate the 
column density with the local density, since the reaction rate is
determined by the local density.}
In the original formulation in FGK12, the dependences on
$n_\mathrm{H}$ and $N_\mathrm{H}$ are both absorbed in
$A_V$ by imposing $A_V\propto n_\mathrm{H}\propto N_\mathrm{H}$.
In this paper, the proportionality constant between $A_V$ and $N_\mathrm{H}$
is given consistently with our model, while we use the relation given
in FGK12 for $A'_V$ and $N'_\mathrm{H}$ (see below), {where
we put a prime to the quantities
calculated in GM11 to distinguish them from the quantities calculated in our model.}
We always adopt $N_\mathrm{H}/n_\mathrm{H}=N'_\mathrm{H}/n'_\mathrm{H}$
to make the following equations solvable (following FGK12).

The CO abundance $x_\mathrm{CO}$ is defined as the number
ratio of CO molecules to hydrogen nuclei. We assume, following FGK12,
that the CO abundance
is determined by the dust extinction and the radiation field
under a given metallicity; thus,
we write the CO abundance as
$x_\mathrm{CO}=x_\mathrm{CO}(A_V,\, \chi,\, Z)$.

{Now let us formulate the estimation method of the CO abundance.}
We utilize the calculation results in GM11 to estimate $x_\mathrm{CO}$.
GM11 provide CO fraction $x'_\mathrm{CO}$  at
$\chi ' =1.7.$\footnote{The Milky Way interstellar radiation field adopted by GM11
is based on \citet{Draine:1978aa}'s estimate, which
corresponds to $\chi =1.7$.
As shown later, we can practically regard $x'_\mathrm{CO}$
as a function of $A'_V$.}

Following FGK12, we start with the equilibrium
between dissociation and formation of CO in the two systems
that realize CO abundances of $x_\mathrm{CO}$ and
$x'_\mathrm{CO}$ (recall that the quantities with a prime mean those obtained
in GM11's simulation):
\begin{align}
x_\mathrm{CO}\chi S_\mathrm{dust}(A_{V,\mathrm{eff}})S_\mathrm{H_2}(N_\mathrm{H_2})
S_\mathrm{CO}(N_\mathrm{CO})=n_\mathrm{H}\sum_{i,j}\mathcal{R}_{i,j}x_ix_j,
\label{eq:equil1}
\end{align}
and
\begin{align}
1.7x'_\mathrm{CO}S_\mathrm{dust}(A'_V)S_\mathrm{H_2}(N'_\mathrm{H_2})
S_\mathrm{CO}(N'_\mathrm{CO})=n'_\mathrm{H}\sum_{i,j}\mathcal{R}_{i,j}x'_ix'_j,
\label{eq:equil2}
\end{align}
where $\mathcal{R}_{i,j}$ is the reaction rates of CO formation from
species $i$ and $j$, the abundances of which are denoted as
$x_i$ and $x_j$ (or $x'_i$ and $x'_j$), respectively.
As mentioned in FGK12,
it is plausible that 
$x_i\sim x'_i$ and $x_j\sim x'_j$ under $x_\mathrm{CO}=x'_\mathrm{CO}$.
Thus, {we expect that the left-hand side of equation (\ref{eq:equil1})
divided by $n_\mathrm{H}$ is equal to the left-hand side of equation (\ref{eq:equil2})
divided by $n'_\mathrm{H}$.}
Recalling that
we adopt $n_\mathrm{H}/n'_\mathrm{H}=N_\mathrm{H}/N'_\mathrm{H}$
and using $x_\mathrm{CO}=x'_\mathrm{CO}$, we obtain
\begin{multline}
\chi S_\mathrm{dust}(A_{V,\mathrm{eff}})S_\mathrm{H_2}(N_\mathrm{H_2})
S_\mathrm{CO}(N_\mathrm{CO})/N_\mathrm{H}\\
=
1.7S_\mathrm{dust}(A'_V)S_\mathrm{H_2}(N'_\mathrm{H_2})
S_\mathrm{CO}(N'_\mathrm{CO})/N'_\mathrm{H},\label{eq:diss_eq}
\end{multline}
where
$S_\mathrm{dust}(A_V)$, $S_\mathrm{H_2}(N_\mathrm{H_2})$, and
$S_\mathrm{CO}(N_\mathrm{CO})$ are the shielding factors of
CO-dissociating photons ($1-S$ is the fraction of shielded CO-dissociating
radiation by each species) by dust, H$_2$, and CO, respectively,
and $A_{V,\mathrm{eff}}=A_\mathrm{1000~\AA}/4.7$
is the effective $V$-band extinction.
The shielding functions $S$'s are taken from \citet{Lee:1996aa}.
To obtain $A_{V,\mathrm{eff}}$, we first obtain
$A_\mathrm{1000~\AA}\simeq 1.086\sum_i\tau_{\mathrm{LW},i}$
using equation (\ref{eq:tau_LW}).
We convert this to $A_{V,\mathrm{eff}}$ with the above relation,
where the factor 4.7 comes from the
Milky Way extinction curve \citep{Pei:1992aa}.

{Equation (\ref{eq:diss_eq}) is the one we solve to
obtain the CO abundance.
Note that all the quantities on the left-hand side in
equation (\ref{eq:diss_eq}) are given
by our simulation except for
$N_\mathrm{CO}=x_\mathrm{CO}N_\mathrm{H}=x'_\mathrm{CO}N_\mathrm{H}$
while those on the right-hand side are given as a function of $A'_V$ as we see
below. Because $x'_\mathrm{CO}$ is a function of $A'_V$ as we see
below (thus, with the assumption of $x_\mathrm{CO}=x'_\mathrm{CO}$,
$x_\mathrm{CO}$ is also a function of $A'_V$),
equation (\ref{eq:diss_eq}) is an equation for $A'_V$. Once we obtain
$A'_V$, we calculate $x_\mathrm{CO}=x'_\mathrm{CO}(A'_V)$.}


{Now we explain how $N'_\mathrm{H_2}$ and $N'_\mathrm{CO}$
are expressed as a function of $A'_V$.}
Based on GM11's result,
we adopt the following fitting formulae:
\begin{align}
f'_\mathrm{H_2}=1-\exp (-0.45A'_V),\label{eq:fH2_Av}
\end{align}
and
\begin{align}
\log x'_\mathrm{CO}=-7.64+3.89\log A'_V,\label{eq:xco_Av}
\end{align}
where the H$_2$ fraction $f'_\mathrm{H_2}$ is related to
$N'_\mathrm{H_2}$ as
$N'_\mathrm{H_2}=f'_\mathrm{H_2}N'_\mathrm{H}/2$,
and
\begin{align}
N'_\mathrm{H}=
\frac{A'_V}{5.348\times 10^{-22}(Z'/\mathrm{Z}_{\sun})~\mathrm{cm}^{2}}
\label{eq:NH_Av}
\end{align}
in GM11's formulation (we adopt $Z'=Z$).
Note that these relations (equation \ref{eq:fH2_Av}--\ref{eq:NH_Av}) do not
hold among $f_\mathrm{H_2}$, $x_\mathrm{CO}$, $N_\mathrm{H}$ and
$A_V$ because physical conditions such as dust properties and
radiation field are different between our models and
GM11's calculations.
{We also use $N_\mathrm{H_2}=f_\mathrm{H_2}N_\mathrm{H}/2$
and $N_\mathrm{CO}=x_\mathrm{CO}N_\mathrm{H}$, and use
$f_\mathrm{H_2}$ calculated in Section \ref{subsec:H2}.
As mentioned above, we are looking for a solution that satisfies
$x_\mathrm{CO}=x'_\mathrm{CO}$. Thus, equation (\ref{eq:diss_eq}) is rewritten as}
\begin{multline}
\chi S_\mathrm{dust}(A_{V,\mathrm{eff}})S_\mathrm{H_2}(f_\mathrm{H_2}N_\mathrm{H}/2)
S_\mathrm{CO}(x'_\mathrm{CO}N_\mathrm{H})/N_\mathrm{H}\\
=
1.7S_\mathrm{dust}(A'_V)
S_\mathrm{H_2}(f'_\mathrm{H_2}N'_\mathrm{H}/2)
S_\mathrm{CO}(x'_\mathrm{CO}N'_\mathrm{H})/N'_\mathrm{H},\label{eq:feldmann}
\end{multline}
where $f'_\mathrm{H_2}$, $x'_\mathrm{CO}$, and
$N'_\mathrm{H}$ are all functions of $A'_V$ through
equations (\ref{eq:fH2_Av}), (\ref{eq:xco_Av}), and
(\ref{eq:NH_Av}), respectively.
Note that $N_\mathrm{H}$ is a given parameter,
and that $A_{V,\mathrm{eff}}$ and $f_\mathrm{H_2}$ are
calculated by the model. Thus, we solve 
equation (\ref{eq:feldmann}) to obtain $A'_V$,
which is translated into the CO abundance
$x'_\mathrm{CO}=x_\mathrm{CO}$ through equation (\ref{eq:xco_Av}).

We impose an upper limit
$x_\mathrm{C}=1.41\times 10^{-4}Z/\mathrm{Z}_{\sun}$
(carbon abundance) for
$x_\mathrm{CO}$; {that is,
if the obtained $x_\mathrm{CO}$ is larger than the carbon abundance,
we adopt $x_\mathrm{CO}=x_\mathrm{C}$. This only occurs at
super-solar metallicities in our models.}

\subsection{CO-to-H$_2$ conversion factor}

Using the quantities calculated above,
we finally calculate the CO-to-H$_2$ conversion
factor:
\begin{align}
X_\mathrm{CO}=N_\mathrm{H_2}/W_\mathrm{CO}.\label{eq:Xco}
\end{align}
We compute $W_\mathrm{CO}$ using the following expression
(GM11):
\begin{align}
W_\mathrm{CO}=T_\mathrm{r}\Delta v\int_0^{\tau_{10}}
2\beta (\tau )\,\mathrm{d}\tau,\label{eq:W_CO}
\end{align}
where $T_\mathrm{r}$ is the observed radiation temperature
(calculated later in equation \ref{eq:temp_rad}),
$\Delta v$ is the velocity width of the CO line,
$\tau_{10}$ is the optical depth of the CO $J=1\to 0$  transition,
and $\beta (\tau )$ is the photon escape probability as a function
of the optical depth. The escape probability is estimated as
\citep{Tielens:2005aa}
\begin{align}
\beta (\tau )= \begin{cases}
[1-\exp (-2.34\tau )]/(4.68\tau ) & \mbox{if $\tau\leq 7$}; \\
1/(4\tau [\ln (\tau /\sqrt{\pi})]^{1/2}) & \mbox{if $\tau >7$}.
\end{cases}
\end{align}
The optical depth $\tau_{10}$ is estimated as
(\citealt{Tielens:2005aa}; FGK12)
\begin{align}
\tau_{10}=1.4\times 10^{-16}(1-\mathrm{e}^{-5.5/T_\mathrm{gas}})
\left(\frac{\Delta v}{3~\mathrm{km~s}^{-1}}\right)^{-1}
\left(\frac{N_\mathrm{CO}}{\mathrm{cm}^{-2}}\right) .
\end{align}
The radiation temperature of the CO $J=1\to 0$
transition with the subtraction of the cosmic microwave background
(CMB) taken into account
is calculated by
\begin{align}
T_\mathrm{r}=5.5\left(\frac{1}{\mathrm{e}^{5.5/T_\mathrm{gas}}-1}-
\frac{1}{\mathrm{e}^{5.5/T_\mathrm{CMB}}-1}\right)~\mathrm{K} ,
\label{eq:temp_rad}
\end{align}
where $T_\mathrm{CMB}=2.73(1+z)$ is the CMB temperature
($z$ is the redshift of the galaxy
considered; we adopt $z=0$ in this paper).
For the velocity width, we simply adopt the typical value adopted
in FGK12, $\Delta v=3$ km s$^{-1}$, partly because our models are
based on their model, partly because a direct comparison with their
model is possible (i.e.\ our models are `calibrated' by their model).
Using equation (\ref{eq:W_CO}) and the H$_2$ column density
calculated in Section \ref{subsec:H2}, we finally obtain the
CO-to-H$_2$ conversion factor, $X_\mathrm{CO}$, by
equation (\ref{eq:Xco}).

\subsection{Choice of parameter values}\label{subsec:choice}

For the dust evolution model, we adopt $\mathcal{R} = 0.25$
and $\mathcal{Y}_Z = 0.013$
\citep{Hirashita:2011aa}. For the parameters governing the
dust evolution, we adopt the same ranges as adopted in H15
as listed in Table \ref{tab:param}.

\begin{table}
\centering
\begin{minipage}{80mm}
\caption{Parameter ranges surveyed for dust evolution.}
\label{tab:param}
\begin{center}
\begin{tabular}{@{}llccc} \hline
Parameter & Process & Minimum & Maximum & Fiducial
\\ \hline
$f_\mathrm{in}$ & stellar ejecta & 0.01 & 0.1 & 0.1\\
$\beta_\mathrm{SN}$ & SN destruction & 4.8 & 19 & 9.65\\
$\tau_\mathrm{cl}$ & accretion & $10^6$ yr & $10^8$ yr & $10^7$ yr\\
$\tau_\mathrm{sh,0}$ & shattering & $10^7$ yr & $10^9$ yr & $10^8$ yr\\
$\tau_\mathrm{co,0}$ & coagulation & $10^6$ yr & $10^8$ yr & $10^7$ yr\\
$\tau_\mathrm{SF}$ & star formation & 0.5 Gyr & 50 Gyr & 5 Gyr\\ \hline
\end{tabular}
\end{center}
\end{minipage}
\end{table}

As assumed in Section \ref{subsec:xco}, we give $\chi$ and
$N_\mathrm{H}$ for the physical condition of the cloud at a given
$Z$. Although $n_\mathrm{H}$ is not essential to our formulation
for the CO abundance, it is necessary in calculating the H$_2$
abundance. Whenever necessary, we adopt
$n_\mathrm{H}=10^3$ cm$^{-2}$, which is appropriate for
the molecular clouds in the Milky Way (but not necessarily
`molecular' in galaxies with higher $\chi$ or lower $Z$ than in
the Milky Way) \citep{Hirashita:2011aa}.
For the column density, we investigate a range of
$N_\mathrm{H}=10^{21}$--$10^{23}$ cm$^{-2}$, corresponding
to surface densities of 10--$10^3$ M$_{\sun}$ pc$^{-2}$.
This covers the range of surface densities of giant molecular
clouds in various environments \citep[e.g.][]{Bolatto:2013aa}.
For the UV intensity normalized to the
\citet{Habing:1968aa} value,
\citet{Hirashita:2005aa} derived the relation between
$\chi$ and the surface density of SFR ($\Sigma_\mathrm{SFR}$)
by assuming that the UV luminosity traces the current star formation
rate as
\begin{align}
\Sigma_\mathrm{SFR}=1.7\times 10^{-3}\chi~\mathrm{M}_{\sun}~\mathrm{yr}^{-1}~
\mathrm{kpc}^{-2}.
\end{align}
We survey a range of
$\chi =1.7$--170 (1--100 times the Galactic value),
corresponding to $\Sigma_\mathrm{SFR}=2.9\times 10^{-3}$--0.29
$\mathrm{M}_{\sun}~\mathrm{yr}^{-1}~
\mathrm{kpc}^{-2}$, which roughly covers the range of
nearby disc galaxies \citep{Kennicutt:1998aa}
(see also Section \ref{subsec:various_galaxies}).
Starburst galaxies have $\Sigma_\mathrm{SFR}$
10--100 times higher $\Sigma_\mathrm{SFR}$
\citep{Kennicutt:1998aa}, thus $\chi$.
We consider such extreme values of $\chi$ later in
Section \ref{subsec:starburst}.
In Table \ref{tab:param_cloud}, we list the ranges
and fiducial values of $N_\mathrm{H}$ and $\chi$.

\begin{table}
\centering
\begin{minipage}{70mm}
\caption{Parameter ranges for the physical condition of clouds.}
\label{tab:param_cloud}
\begin{center}
\begin{tabular}{@{}lccc} \hline
Parameter & Minimum & Maximum & Fiducial
\\ \hline
$N_\mathrm{H}$ & $10^{21}$ cm$^{-2}$ & $10^{23}$ cm$^{-3}$ & $10^{22}$ cm$^{-3}$\\
$\chi$ & 1.7 & 170 & 1.7\\ \hline
\end{tabular}
\end{center}
\end{minipage}
\end{table}

\section{Results}\label{sec:result}

\subsection{Basic properties}\label{subsec:basic}

\begin{figure}
 \includegraphics[width=\columnwidth]{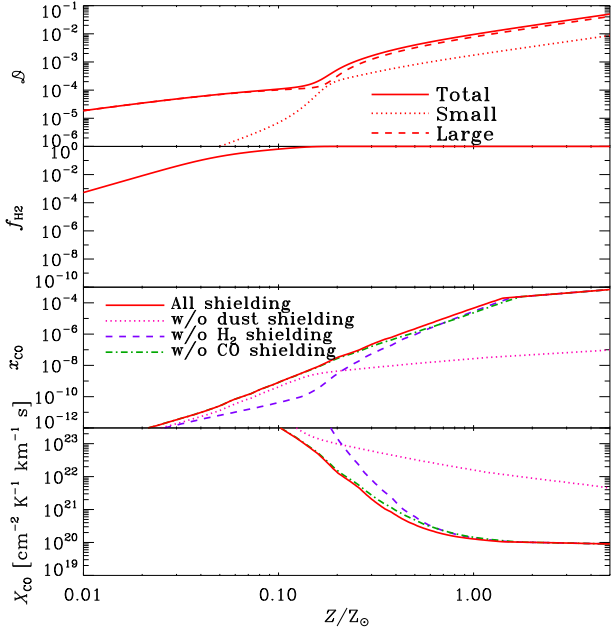}
 \caption{Quantities of interest as a function of metallicity.
 Panels from upper to lower show the dust-to-gas ratio ($\mathcal{D}$),
 H$_2$ fraction ($f_\mathrm{H_2}$), CO fraction ($x_\mathrm{CO}$)
 and CO-to-H$_2$ conversion factor ($X_\mathrm{CO}$).
 For the dust-to-gas ratio, the small and large grain components
 are shown by the dotted and dashed lines, respectively, and the
 total dust-to-gas ratio is shown by the solid line.
 For $x_\mathrm{CO}$ and $X_\mathrm{CO}$, the cases without either
 dust shielding (dotted),
 H$_2$ shielding (dashed), or CO self-shielding (dot-dashed) are also shown.
 $X_\mathrm{CO}$ is shown only when $x_\mathrm{CO}>10^{-10}$.}
 \label{fig:fiducial}
\end{figure}

Here we investigate the general evolution of H$_2$ and CO abundances
under various conditions. We adopt the fiducial values for the parameters
of dust evolution and cloud properties (Tables \ref{tab:param} and
\ref{tab:param_cloud}).
In Fig.\ \ref{fig:fiducial}, we show the metallicity dependence of  the quantities
investigated in this paper,
namely, the dust-to-gas ratio ($\mathcal{D}$), 
H$_2$ fraction ($f_\mathrm{H_2}$), CO fraction ($x_\mathrm{CO}$)
and CO-to-H$_2$ conversion factor ($X_\mathrm{CO}$).
$X_\mathrm{CO}$ is only shown where $x_\mathrm{CO}>10^{-10}$,
below which detection of CO is impossible in any case.

H15 already investigated and discussed the $\mathcal{D}$--$Z$
relation, which is briefly described in what follows.
Dust is supplied by stars at low metallicities, and is dominated
by large grains. Small grains gradually increase because of
shattering of large grains. At a certain point,
the increase of small grains is accelerated because of
accretion, which induces the increase of large grains as well
through coagulation. As a consequence, the total dust
abundance steeply increases as seen around $Z\sim 0.1$--0.2 Z$_{\sun}$.
After that, the dust-to-gas ratio gradually increases because of the
metal enrichment; at this stage, the dust-to-gas ratio is determined
by the balance between accretion and SN destruction.

The H$_2$ fraction, $f_\mathrm{H_2}$ also increases as the
dust-to-gas ratio increases.
A fully molecular condition is realized at $Z\ga 0.1$ Z$_{\sun}$.
Thus, a cloud with a typical density and column density of
Milky Way `molecular' clouds is not fully molecular below
metallicity 0.1~Z$_{\sun}$.
The CO fraction, $x_\mathrm{CO}$, also increases
associated with the increase of dust-to-gas ratio.
To understand what drives the increase of $x_\mathrm{CO}$, we
also show the cases without either dust shielding, H$_2$ shielding,
or CO self-shielding in the third window in Fig.\ \ref{fig:fiducial}.
The prediction without H$_2$ shielding underproduces the CO
abundance at low metallicity $\la 0.2$ Z$_{\sun}$, which indicates
that H$_2$ shielding is important for CO formation. This is
associated with the metallicity at which the cloud becomes fully
molecular. At $Z\ga 0.2$ Z$_{\sun}$, the main shielding component
changes to dust: If we do not include dust shielding, the
CO abundance remains as low as $\sim 10^{-6}$--$10^{-5}$.
Dust shielding becomes important after the dust abundance
is increased by accretion. Self-shielding of CO has an influence
on the CO abundance only around solar metallicity, but
its contribution is minor.
The CO abundance, $x_\mathrm{CO}$ cannot be
larger than the carbon abundance ($x_\mathrm{C}$);
thus, at $Z\ga 2$ Z$_{\sun}$,
$x_\mathrm{CO}$ is limited by the carbon abundance.

The CO-to-H$_2$ conversion factor, $X_\mathrm{CO}$ is
sensitive to metallicity. This is particularly because there is a metallicity
range where the cloud is rich in H$_2$ but is not rich in
CO at $Z\ga 0.1$ Z$_{\sun}$. Around solar metallicity,
$X_\mathrm{CO}$ is near to the value observed in the Milky Way
($\sim 2\times 10^{20}$ cm$^{-2}$ K$^{-1}$ km$^{-1}$ s;
\citealt{Bolatto:2013aa}). Above solar metallicity,
$X_\mathrm{CO}$ is not sensitive to metallicity because
the CO $J=1\to 0$ line is optically thick. We also show the metallicity
dependence of the CO-to-H$_2$ conversion factor without
one of the shielding mechanisms. H$_2$ shielding has an influence on
the slope of the $X_\mathrm{CO}$--$Z$ relation, since it has
a larger impact on the CO abundance at low metallicity as discussed
above. Dust shielding has a dramatic impact: $X_\mathrm{CO}$
is two orders of magnitude higher than the Milky way value
at solar metallicity {if we do not include dust shielding}.
CO self-shielding has little influence on $X_\mathrm{CO}$.

\subsection{Dependence on the cloud parameters}

\begin{figure*}
 \includegraphics[width=0.98\columnwidth]{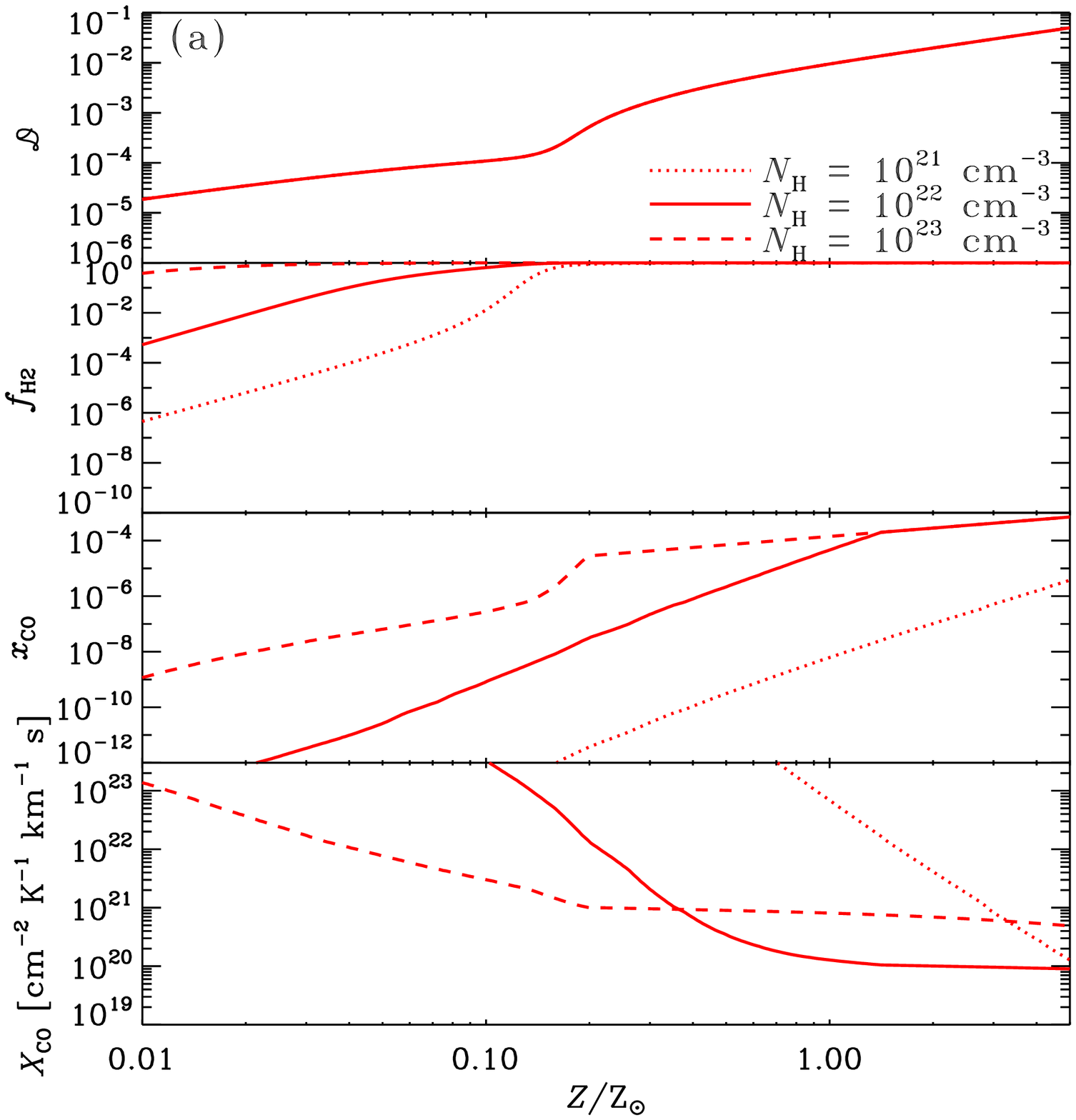}
 \includegraphics[width=0.98\columnwidth]{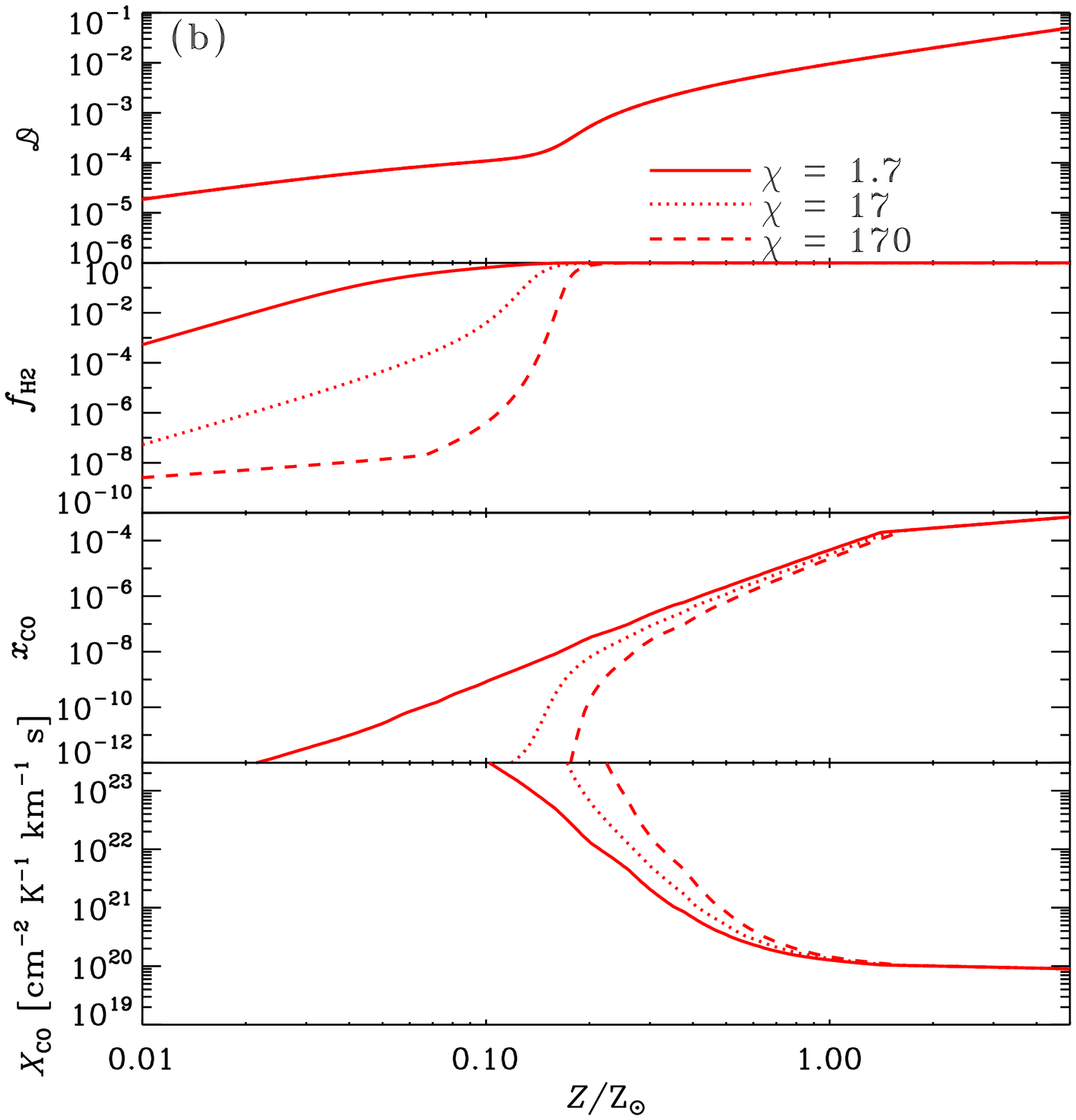}
 \caption{Same as Fig.\ \ref{fig:fiducial} but with the following parameters
 changed:
 (a) {$N_\mathrm{H}=10^{21}$, $10^{22}$, and $10^{23}$ cm$^{-2}$
 for the dotted, solid, and dashed lines, respectively, with $\chi =1.7$.
 (The solid line always shows the fiducial case in all the figures in this paper.)}
 (b) $\chi =1.7$, 17, and 170 for the solid, dotted, and dashed lines,
 respectively, with $N_\mathrm{H}=10^{22}$ cm$^{-2}$.
 We fix the values of
 the parameters other than the changed parameter at the fiducial values listed
 in Table \ref{tab:param_cloud}.
 We adopt the dust evolution model characterized by the fiducial parameters
 in Table \ref{tab:param}.}
 \label{fig:dependence_nH}
\end{figure*}

We investigate the dependence on the cloud parameters
listed in Table \ref{tab:param_cloud} (i.e.\
$N_\mathrm{H}$ and $\chi$). In Fig.\ \ref{fig:dependence_nH},
we show the variation of the quantities of interest as a function
of metallicity. Note that the $\mathcal{D}$--$Z$ relation is not
affected by the change of those parameters.

In Fig.\ \ref{fig:dependence_nH}a, we show the dependence on
the column density ($N_\mathrm{H}$). Because of stronger
self-shielding, $f_\mathrm{H_2}$ is larger for a larger $N_\mathrm{H}$
($f_\mathrm{H_2}\propto N_\mathrm{H}^3$ for
the self-shielded regime as long as $f_\mathrm{H_2}\ll 1$;
\citealt{Hirashita:2005aa}).
The increase of dust abundance around $Z\sim 0.1$~Z$_{\sun}$
is important to raise the H$_2$ fraction for $N_\mathrm{H}=10^{21}$~cm$^{-2}$;
thus, the increase of dust abundance by accretion is important to make
the cloud molecule-rich at low column densities.
The CO abundance is very sensitive to
$N_\mathrm{H}$; at $Z\la 0.1$~Z$_\odot$, the CO abundance is
governed by H$_2$ shielding, while at higher metallicities,
it is regulated by dust shielding (Section \ref{subsec:basic}).
In the case of
$N_\mathrm{H}=10^{23}$~cm$^{-2}$, the rapid change of
$x_\mathrm{CO}$ around $Z\sim 0.1$ Z$_{\sun}$ is due to the
rapid increase of dust abundance (or dust shielding).
The CO-to-H$_2$ conversion factor, $X_\mathrm{CO}$,
drops to $10^{21}$ cm$^{-2}$ K$^{-1}$ km$^{-1}$ s even at
$Z\sim 0.2$ Z$_{\sun}$ in the case of $N_\mathrm{H}=10^{23}$ cm$^{-2}$,
while it does not drop further because of high CO optical depth.
In contrast, $X_\mathrm{CO}$ for $N_\mathrm{H}=10^{21}$ cm$^{-2}$
is much higher than the
Galactic value ($\sim 2\times 10^{20}$ cm$^{-2}$ K$^{-1}$ km$^{-1}$ s)
because of low dust shielding.
Around solar metallicity, the case with $N_\mathrm{H}=10^{22}$ cm$^{-2}$
has the smallest $X_\mathrm{CO}$ {among the three column densities}.
This is consistent with the theoretical expectation by
FGK12 (see their figure 3): as mentioned above,
at $N_\mathrm{H}\la 10^{22}$ cm$^{-2}$, $X_\mathrm{CO}$ is
dominated by dust shielding so that $X_\mathrm{CO}$ is lower
at higher $N_\mathrm{H}$, while at $N_\mathrm{H}\ga 10^{22}$ cm$^{-2}$,
the CO emission is saturated by its large optical depth so that
$X_\mathrm{CO}$ rises for higher $N_\mathrm{H}$. In other words,
the clouds with a typical column density of molecular clouds,
$N_\mathrm{H}\sim 10^{22}$ cm$^{-2}$, is the most efficient in CO
emission per H$_2$ molecule {around solar metallicity}.

In Fig.\ \ref{fig:dependence_nH}b, we present the dependence on the
UV radiation field intensity ($\chi$). In the most intense radiation
($\chi =170$), the H$_2$ fraction is strongly suppressed at a level
where self-shielding is not important ($f_\mathrm{H_2}<2\times 10^{-8}$
for $N_\mathrm{H}=10^{22}$~cm$^{-2}$) at $Z\la 0.07$ Z$_{\sun}$;
at higher metallicities, self-shielding coupled with the drastic increase
of dust abundance makes the sharp transition toward the fully
molecular phase in a narrow metallicity range of 0.007--0.2~Z$_{\sun}$.
The CO fraction is sensitive to $\chi$ at low metallicities where
H$_2$ shielding is important, while it is less sensitive to $\chi$
at $Z\ga 0.3$ Z$_{\sun}$, where
dust shielding is important. This is because
dust shielding, which has an exponential dependence on
the dust-to-gas ratio, sufficiently suppresses the dissociating photons
at high metallicity.
Accordingly, the slope of $X_\mathrm{CO}$ as a function of $Z$
changes in different $\chi$. We will revisit this dependence on
$\chi$ in Section \ref{subsec:starburst}.

\subsection{Effects of the dust evolution parameters}

\begin{figure*}
 \includegraphics[width=0.78\columnwidth]{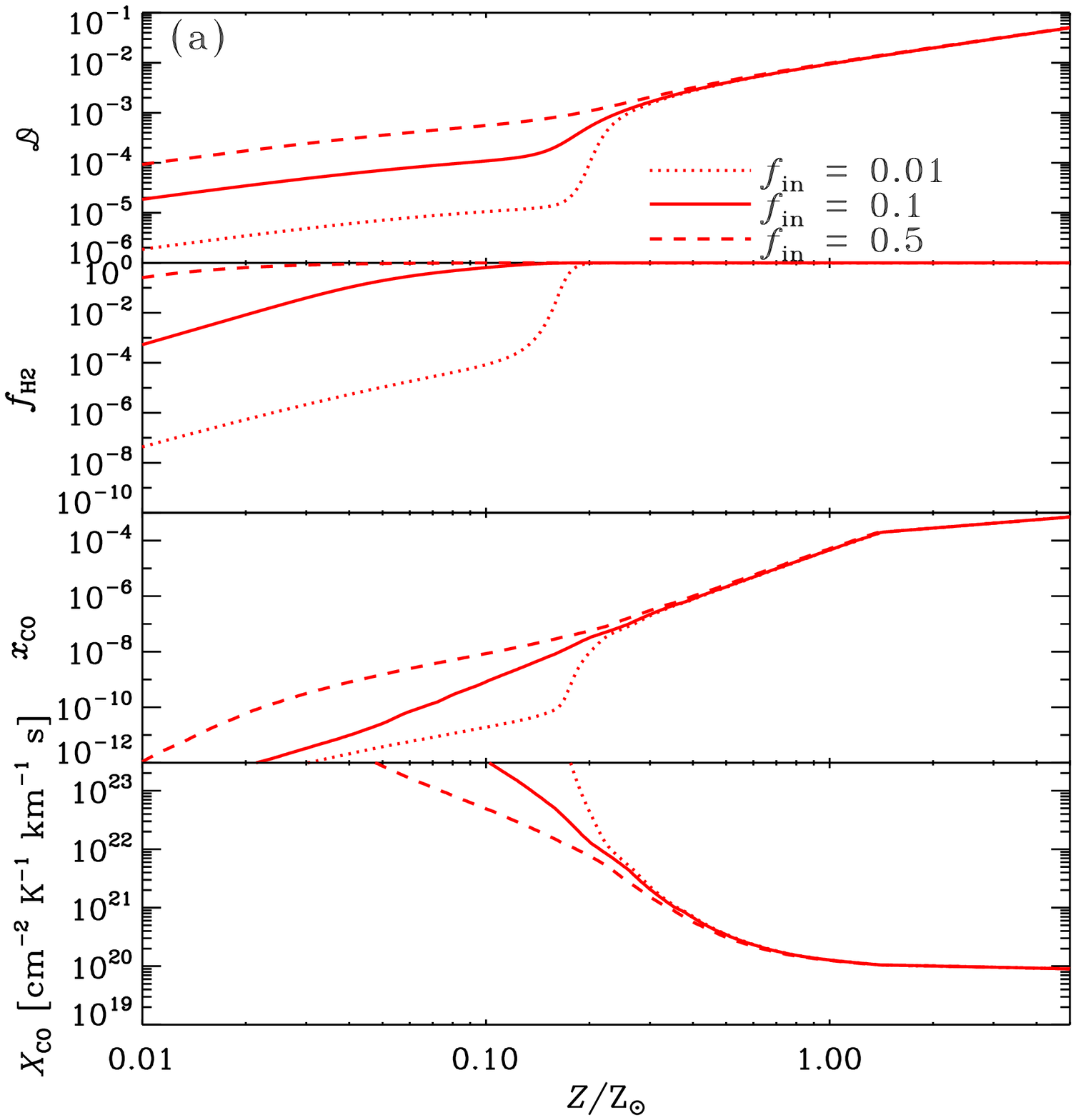}
 \includegraphics[width=0.78\columnwidth]{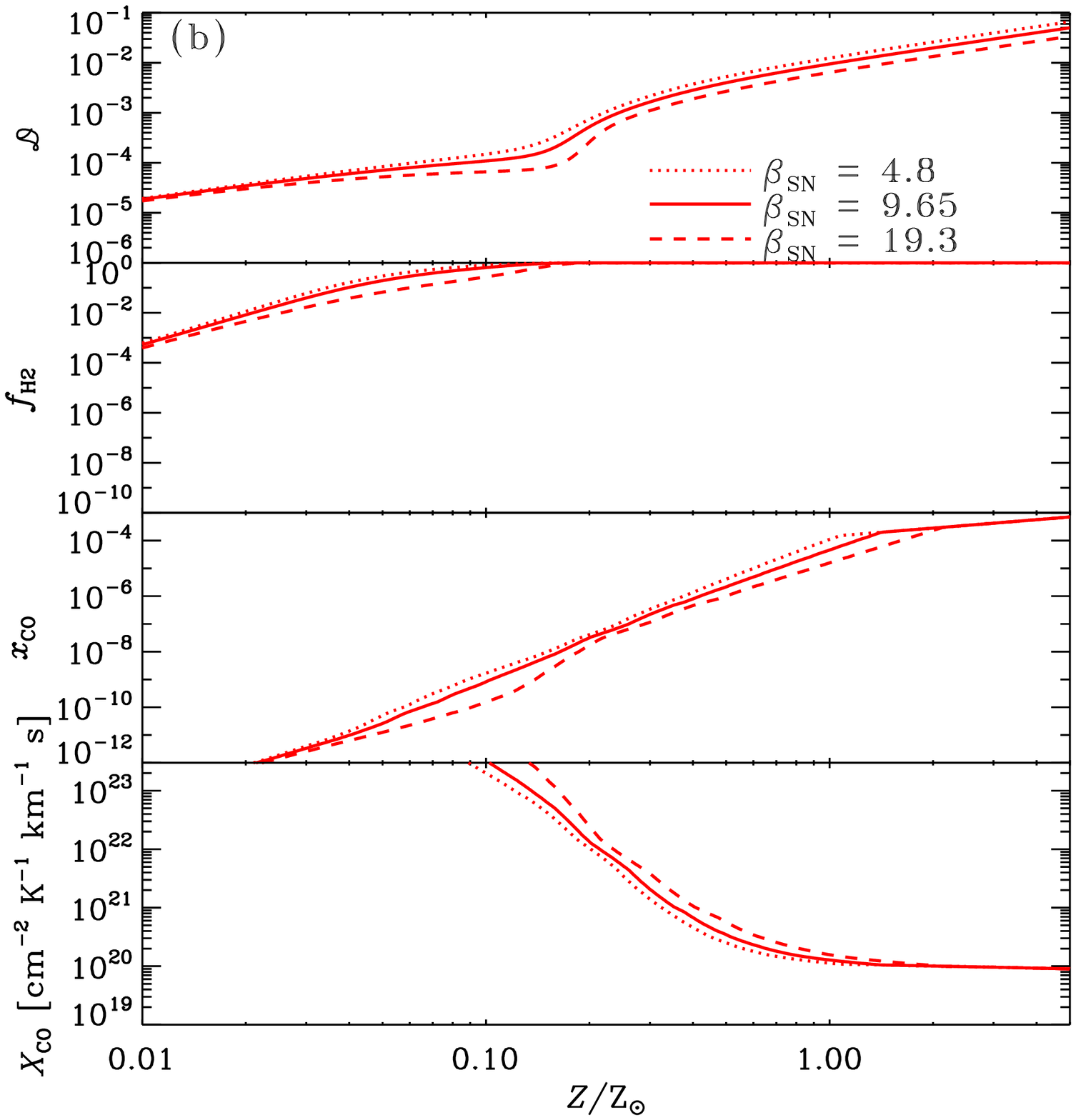}
 \includegraphics[width=0.78\columnwidth]{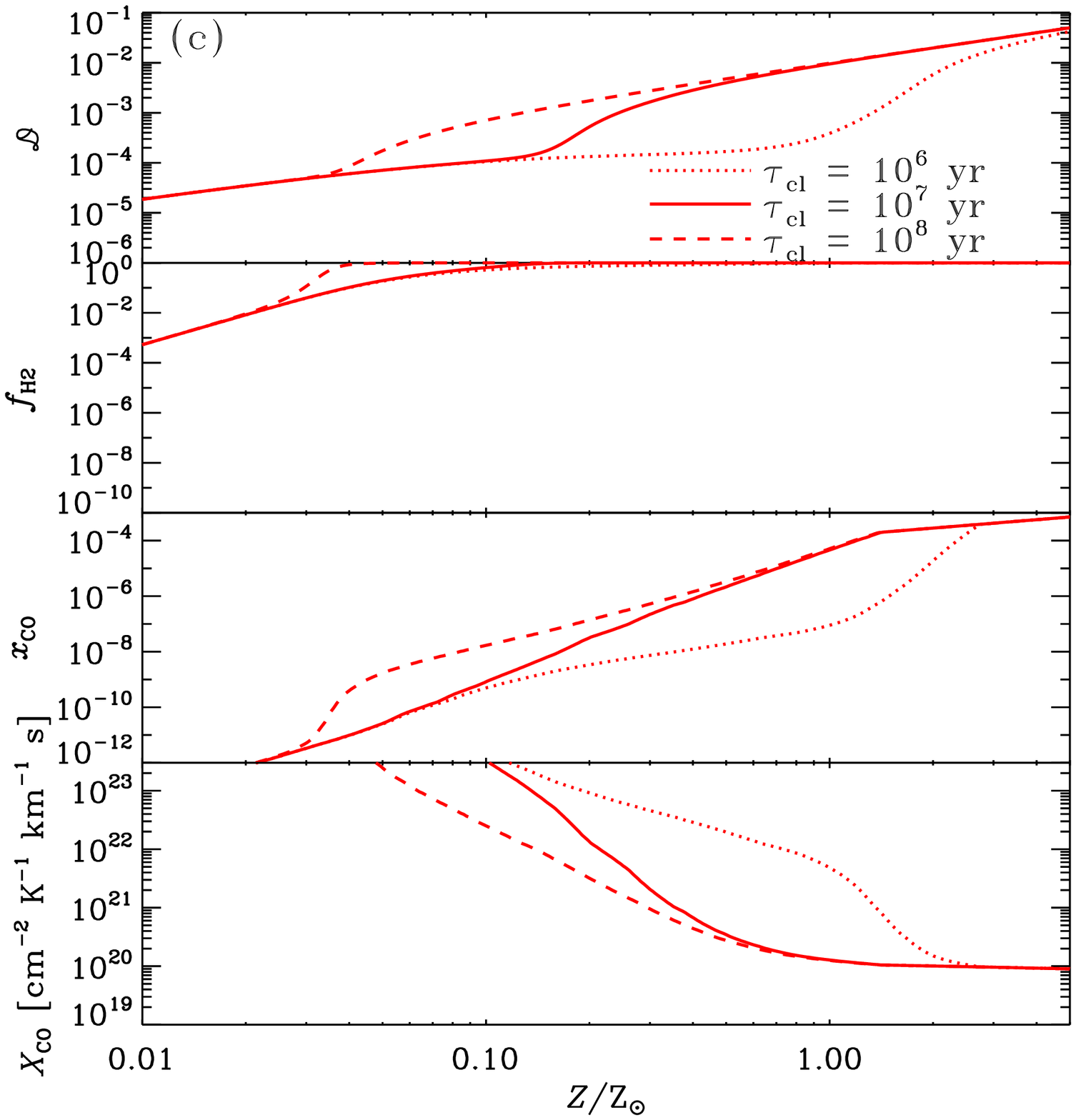}
 \includegraphics[width=0.78\columnwidth]{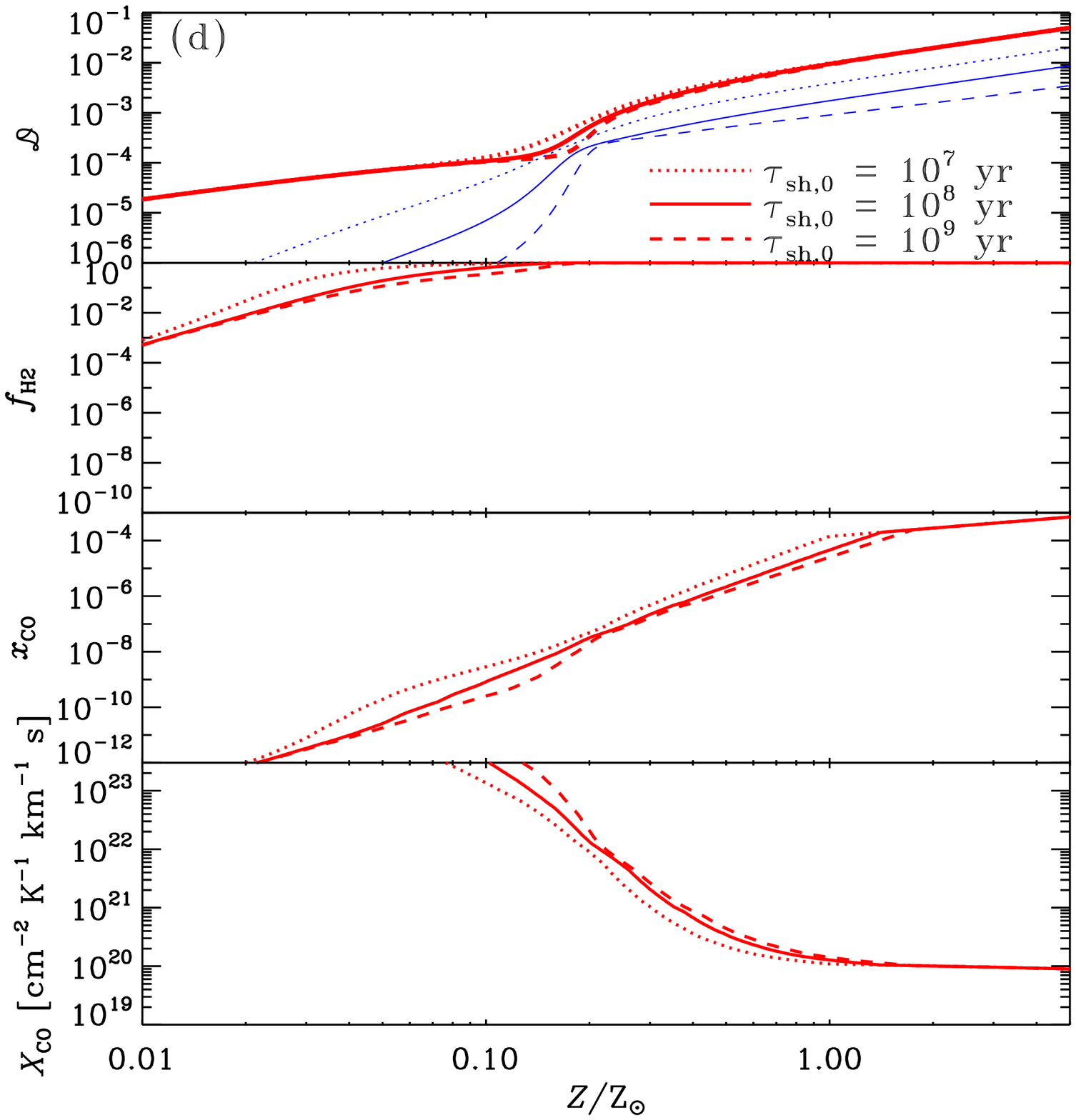}
 \includegraphics[width=0.78\columnwidth]{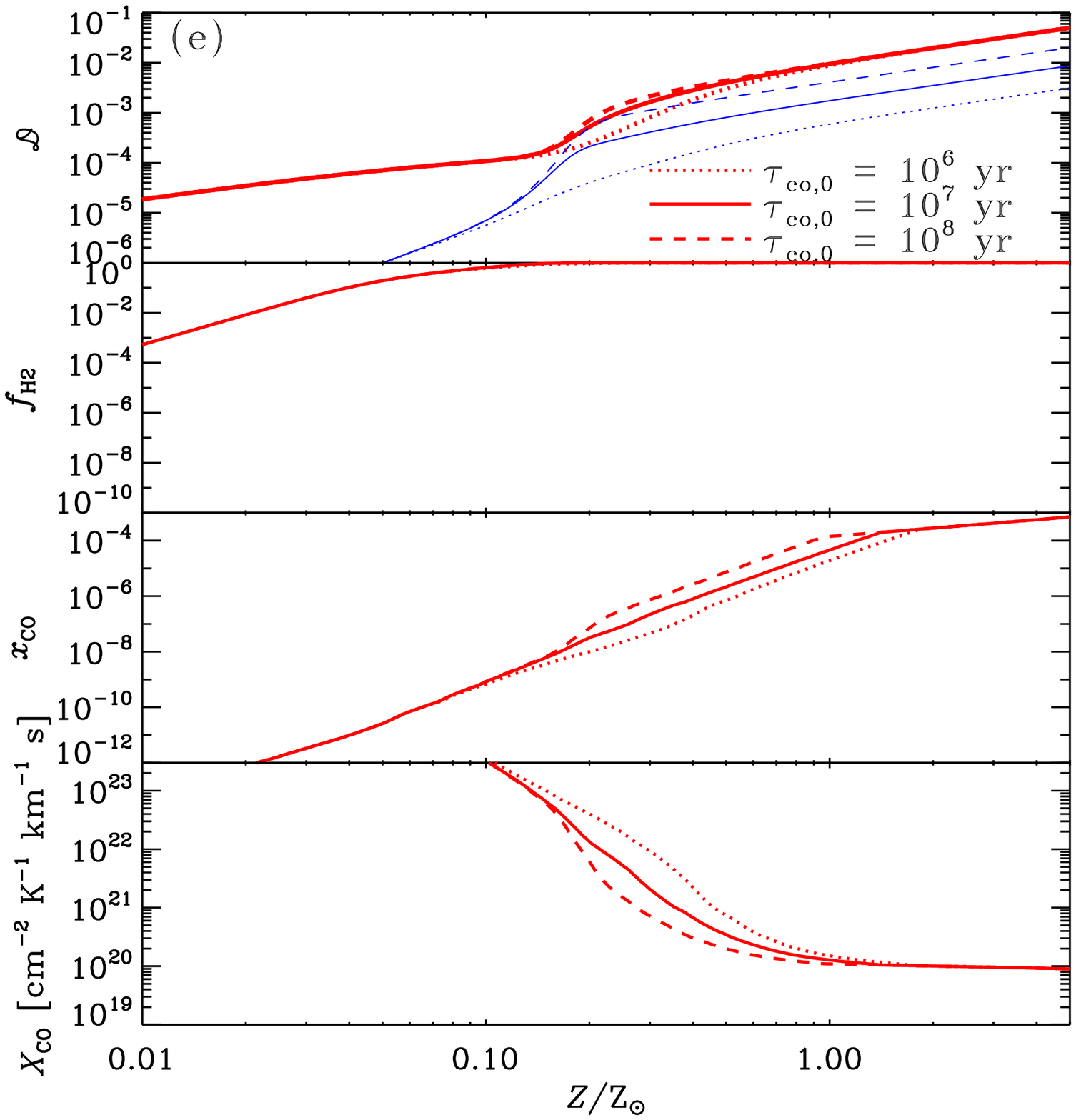}
 \includegraphics[width=0.78\columnwidth]{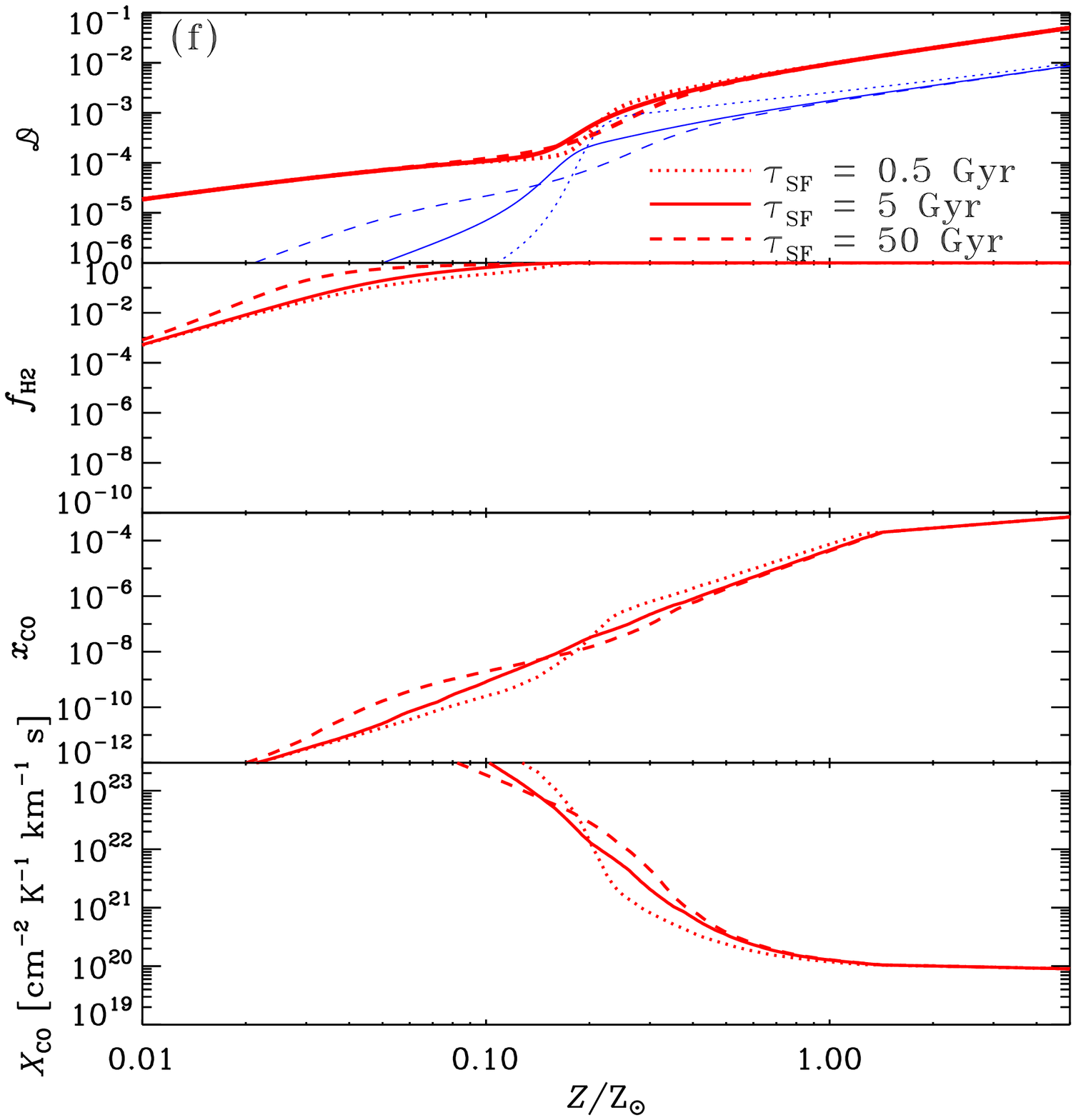}
 \caption{Same as Fig.\ \ref{fig:fiducial} but with the following parameters
 changed: (a) $f_\mathrm{in}=0.01$, 0.1, and 0.5
 for the dotted, solid, and dashed lines, respectively.
 (b) $\beta_\mathrm{SN}=4.8$, 9.65, and 19
 for the dotted, solid, and dashed lines, respectively.
 (c) $\tau_\mathrm{cl} =10^6$, $10^7$, and $10^8$ yr for the dotted, solid, and dashed lines,
 respectively.
 (d) $\tau_\mathrm{sh,0}=10^7$, $10^8$, and $10^9$ yr for the dotted, solid,
 and dashed lines, respectively. In the top window in this panel, we show the total dust-to-gas ratio
 (thick lines) and the small grain dust-to-gas ratio $\mathcal{D}_\mathrm{s}$
 (thin lines), since the effect of shattering is clearly seen in the small-to-large
 grain abundance ratio.
 (e) $\tau_\mathrm{co,0}=10^6$, $10^7$, and $10^8$ yr for the dotted, solid,
 and dashed lines, respectively. In the top window of this panel, we show the total dust-to-gas ratio
 (thick lines) and the small grain dust-to-gas ratio $\mathcal{D}_\mathrm{s}$
 (thin lines), since the effect of coagulation is clear in the small-to-large
 grain abundance ratio.
 (f) $\tau_\mathrm{SF}=0.5$, 5, and 50 Gyr for the dotted, solid, and
 dashed lines, respectively. In the top window of this panel, we show the total dust-to-gas ratio
 (thick lines) and the small grain dust-to-gas ratio $\mathcal{D}_\mathrm{s}$
 (thin lines).
 We fix the values of
 the parameters other than the varied parameter at the fiducial values listed in Table \ref{tab:param}.
 We adopt the fiducial values for the parameters of the cloud in Table \ref{tab:param_cloud}.}
 \label{fig:dependence_dust}
\end{figure*}

We investigate the dependence on the parameters that regulate
the dust evolution (Table \ref{tab:param}). The
resulting variation of the quantities of interest by the change of
the parameters is shown in
Fig.\ \ref{fig:dependence_dust}.

We show the effect of the dust condensation efficiency in stellar ejecta,
$f_\mathrm{in}$ in
Fig.\ \ref{fig:dependence_dust}a. As explained in H15,
the stellar sources of dust dominate the dust-to-gas ratio
at low metallicity before grain growth by accretion becomes
efficient. Thus, the difference in $\mathcal{D}$ between various
$f_\mathrm{in}$ appears at $Z\la 0.2$ Z$_{\sun}$.
As a consequence, the effects of $f_\mathrm{in}$ on
$f_\mathrm{H_2}$ and $x_\mathrm{CO}$ appear at
low metallicity, where
the CO abundance is too low for detection.
Thus, the effect of $f_\mathrm{in}$ on
$x_\mathrm{CO}$ and $X_\mathrm{CO}$ is difficult to be confirmed
observationally. On the other hand, the effect of $f_\mathrm{in}$ on
$f_\mathrm{H_2}$ could be examined if we observe objects with
$Z\la 0.2$ Z$_{\sun}$. A direct determination of H$_2$ abundance is
possible for damped Lyman $\alpha$ objects (DLAs)
using H$_2$ LW absorption lines \citep[e.g.][]{Ledoux:2003aa}.

In Fig.\ \ref{fig:dependence_dust}b, we show the dependence on
$\beta_\mathrm{SN}$, which regulates the efficiency of dust destruction
by SNe.
As expected, a stronger/weaker dust destruction
decreases/increases both $f_\mathrm{H_2}$ and $x_\mathrm{CO}$
because of less/more shielding of dissociating photons.
Around solar metallicity, since the CO abundance is governed
by dust shielding, $x_\mathrm{CO}$ is almost inversely proportional
to $\beta_\mathrm{SN}$. However, $X_\mathrm{CO}$ does not
vary as expected from the change of $x_\mathrm{CO}$ at solar
metallicity, because the CO emission is optically thick.

In Fig.\ \ref{fig:dependence_dust}c, we present the dependence on
$\tau_\mathrm{cl}$, which governs the efficiency of dust growth by
accretion. This parameter determines the metallicity at which
accretion significantly raise the dust-to-gas ratio \citep{Hirashita:2011aa}.
Accretion affects the H$_2$ abundance only if it occurs at
low metallicity such as in the case of $\tau_\mathrm{cl}=10^8$ yr.
Since the increase of CO abundance is associated with the
increase of $\mathcal{D}$, accretion largely affects
$x_\mathrm{CO}$ and $X_\mathrm{CO}$. In particular,
our results predict that the $X_\mathrm{CO}$--$Z$ relation
is sensitive to the metallicity level at which accretion dominates the
increase of dust abundance. In the
case of $\tau_\mathrm{cl}=10^6$~yr (the least efficient accretion),
the effect of accretion only
appears at super-solar metallicity; accordingly, $x_\mathrm{CO}$
remains low and $X_\mathrm{CO}$ is much higher than the
Milky Way value ($\simeq 2\times 10^{20}$ cm$^{-2}$ K$^{-1}$ km$^{-1}$ s)
even at solar metallicity. In the case
of $\tau_\mathrm{cl}=10^8$ yr (the most efficient accretion), in
contrast, a high value of $x_\mathrm{CO}$ at low metallicity
leads to
a shallower slope in the $X_\mathrm{CO}$--$Z$ relation than in the other
cases. Thus, the $X_\mathrm{CO}$--$Z$ relation depends largely on
the efficiency of dust growth by accretion.

In Fig.\ \ref{fig:dependence_dust}d, we show the dependence on
$\tau_\mathrm{sh,0}$, which regulates the efficiency of shattering
(i.e.\ production of small grains from large grains).
Small grains produced by shattering eventually grow by
accretion. Thus, the increase of dust abundance by accretion
appears at the lowest metallicity for the shortest $\tau_\mathrm{sh,0}$.
In the top window of Fig.\ \ref{fig:dependence_dust}d,
we also show the contribution from small grains to the dust-to-gas ratio:
there is a clear difference in the small grain dust-to-gas ratio at all metallicities.
Because small grains have a larger H$_2$ formation rate and
dust extinction per dust mass than large grains, a larger fraction of
small grains (or a higher efficiency of shattering) leads to a higher
molecular abundance. Nevertheless, the effect of shattering on $X_\mathrm{CO}$ is
minor compared with that of accretion.

In Fig.\ \ref{fig:dependence_dust}e, we show the dependence on
the coagulation time-scale, $\tau_\mathrm{co,0}$.
The effect of coagulation becomes prominent after the abundance
of small grains has increased by shattering and accretion; thus, the variation of
coagulation time-scale affects the quantities of interest only
at $Z\ga 0.1$ Z$_{\sun}$.
Because the cloud becomes fully molecular at this metallicity,
coagulation does not have any appreciable impact on the
H$_2$ abundance. Since strong
coagulation suppresses the abundance of small grains, it also
suppresses accretion (recall that the role of accretion is to
increase the small grain abundance; Section \ref{subsec:dust}).
As a result, the total dust abundance
is the smallest in the case of the
most efficient coagulation ($\tau_\mathrm{co,0}=10^6$ yr)
around $Z\sim 0.2$--0.3~Z$_{\sun}$.
Moreover, coagulation also suppresses the relative abundance of
small grains, which leads to a decrease of shielding effect
(recall that small grains have larger dust optical depth per mass).
As a consequence, the CO abundance is lower, and
the CO-to-H$_2$ conversion factor is higher for a shorter
$\tau_\mathrm{co,0}$.

In Fig.\ \ref{fig:dependence_dust}f, we examine the dependence on
the star formation time-scale of the galaxy, $\tau_\mathrm{SF}$.
The star formation time-scale regulates the time-scale of
metal/dust enrichment by stellar sources.
A quick metal enrichment leads to a quick increase in the large-grain
abundance, while the time-scale of shattering is fixed so the production
efficiency of small grains does not increase. Thus, the abundance
of small grains relative to that of large grains is suppressed if
$\tau_\mathrm{SF}$ is short.
Since small grains have larger shielding and H$_2$ formation rate
per mass, the H$_2$ fraction is larger for a longer
$\tau_\mathrm{SF}$.
Because the shattering efficiency is higher if the abundance of large grains is higher,
shattering, if it starts at a high metallicity (i.e.\ large $\mathcal{D}_\mathrm{l}$)
such as in the case of
$\tau_\mathrm{SF}=0.5$ Gyr, has a dramatic impact on the increase
in $\mathcal{D}_\mathrm{s}$. Thus, the increase of $\mathcal{D}_\mathrm{s}$
occurs in a narrow metallicity range for a short $\tau_\mathrm{SF}$.
This produces a
drastic change of $x_\mathrm{CO}$ and $X_\mathrm{CO}$ around
$Z\sim 0.2$--0.3 Z$_{\sun}$.
Therefore, the star formation time-scale
also affects the metallicity dependence of CO abundance in such a way
that mild star formation activities with long $\tau_\mathrm{SF}$
tend to produce smooth increase (decrease) in $x_\mathrm{CO}$
($X_\mathrm{CO}$) as a function of metallicity.

\section{Discussion}\label{sec:discussion}

\subsection{Metallicity dependence of CO-to-H$_2$ conversion factor}
\label{subsec:Zdep}

The main purpose of this work is to clarify how the metallicity
dependence of the H$_2$ and CO abundances is affected by the
dust enrichment and evolution in galaxies.
In particular, the CO-to-H$_2$ conversion factor
has been measured for various types of galaxies with
different metallicities. Indeed, it has been known observationally
that the conversion factor depends strongly on metallicity.
As shown above, the evolution of dust content, which regulates the
abundances of H$_2$ and CO, is also driven by
metallicity. For comparison with observations,
we examine the relation between CO-to-H$_2$ conversion factor and
metallicity.

We adopt the following data sets compiled in \citet{Bolatto:2013aa}.
\citet{Leroy:2011aa} estimated the CO-to-H$_2$ conversion factor
in five Local Group galaxies using dust as a tracer of gas with
a variation of dust-to-gas ratio among the galaxies taken into
account. For their metallicity data given in the form of $12+\log\mathrm{(O/H)}$,
we put an uncertainty of 0.2, and assume that the solar metallicity
corresponds to 8.7 following \citet{Bolatto:2013aa} (we use the
same oxygen abundance for the solar metallicity unless otherwise
stated). For M31, M33, and the Small Magellanic Cloud, we
adopt their spatially resolved data.
\citet{Bolatto:2008aa} derived the CO-to-H$_2$ conversion factor
in nearby galaxies based on virial mass estimates of giant molecular
clouds. We assume the same metallicity uncertainty (0.2) as
assumed in the above sample.
\citet{Israel:1997aa} estimated the CO-to-H$_2$ conversion factor
in the Magellanic Clouds and nearby irregular galaxies. We also
adopted their metallicity data.
\citet{Sandstrom:2013aa} analyzed spatially resolved CO,
dust, and H \textsc{i} maps
on $\sim$kpc scales of nearby star-forming galaxies, and
derived the spatially resolved CO-to-H$_2$ conversion factor.
We adopt their mean CO-to-H$_2$ conversion factor for each galaxy.
We take the metallicity data of their sample from
\citet{Moustakas:2010aa}: among their metallicity calibrations,
we adopt the one by \citet{Pilyugin:2005aa} and assume,
following \citet{Bolatto:2013aa}, that
the solar metallicity corresponds to $12+\log\mathrm{(O/H)}=8.5$.
Besides, we newly add the data taken from
\citet{Cormier:2014aa} for nearby low-metallicity galaxies.
They obtained $X_\mathrm{CO}$, assuming dust to trace the
molecular gas with a variation dust-to-gas ratio among galaxies
taken into account.

{There are two remarks on the observationally derived
CO-to-H$_2$ conversion factors.
The dust-based $X_\mathrm{CO}$ estimates
(i.e.\ other than \citealt{Bolatto:2008aa} above) commonly assume
that the dust-to-gas ratio in molecular clouds is the same as that
in diffuse gas. In reality, we expect that the dust-to-gas ratio
in molecular clouds is higher because of dust growth. If this is true,
the dust-based $X_\mathrm{CO}$ estimate adopts an underestimated dust-to-gas ratio.
A lower dust-to-gas ratio would
overestimate the molecular gas mass by a factor of $\sim 2$
\citep{Leroy:2011aa}, which leads to overestimation
of $X_\mathrm{CO}$ in the observational data.
Since we consider a large dynamic range for $X_\mathrm{CO}$
and there is an order-of-magnitude scatter in the observational data,
a factor 2 difference in dust-to-gas ratio would not affect the overall
comparison shown in this section.
Our theoretical model
also does not distinguish between dust components in molecular clouds
and in diffuse gas, which leads to underestimation of the dust-to-gas ratio
(i.e.\ underestimation of the shielding effect of dissociating photons) in
molecular clouds, and thus, overestimation of $X_\mathrm{CO}$ in
our theoretical models. Thus, if we could consider the difference between
the dust-to-gas ratios between dense and diffuse gas, both theoretical
predictions and observational data of $X_\mathrm{CO}$ would move
in the same direction; thus, it is not clear if the comparison is drastically
affected by dust growth in molecular clouds.
In fact, dust growth could already occur in the diffuse (but relatively
dense) ISM as shown in the numerical simulation by \citet{Zhukovska:2016aa}.
Thus, to fully address the effect of dust growth on $X_\mathrm{CO}$
estimate, we need to use a simulation that includes consistently the evolution of the ISM
and that of dust \citep{Aoyama:2016aa}, and need to include molecule formation
and destruction consistently with the simulation. Such a complete framework is left for future work.}

\begin{figure*}
 \includegraphics[width=0.85\columnwidth]{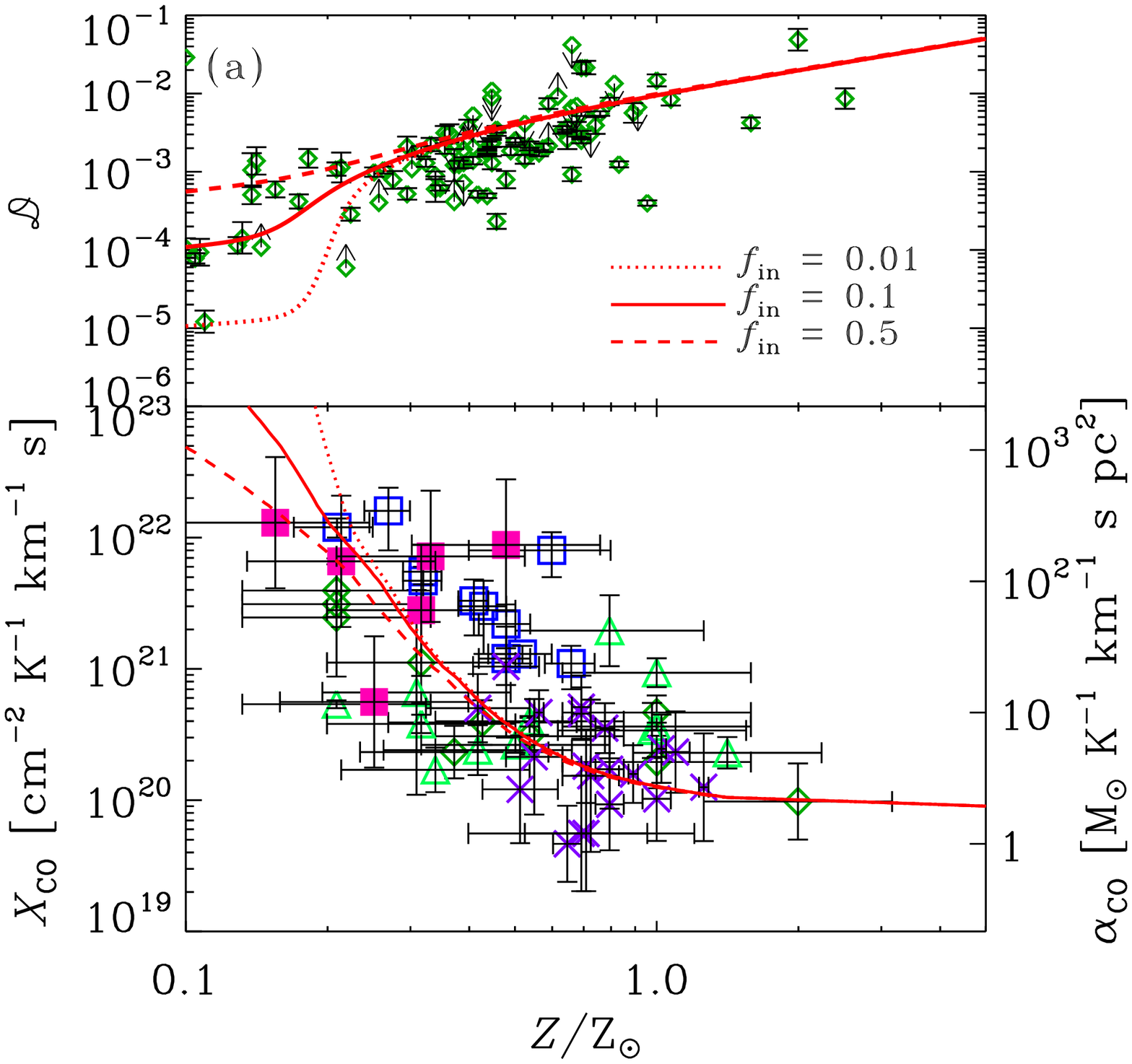}
 \includegraphics[width=0.85\columnwidth]{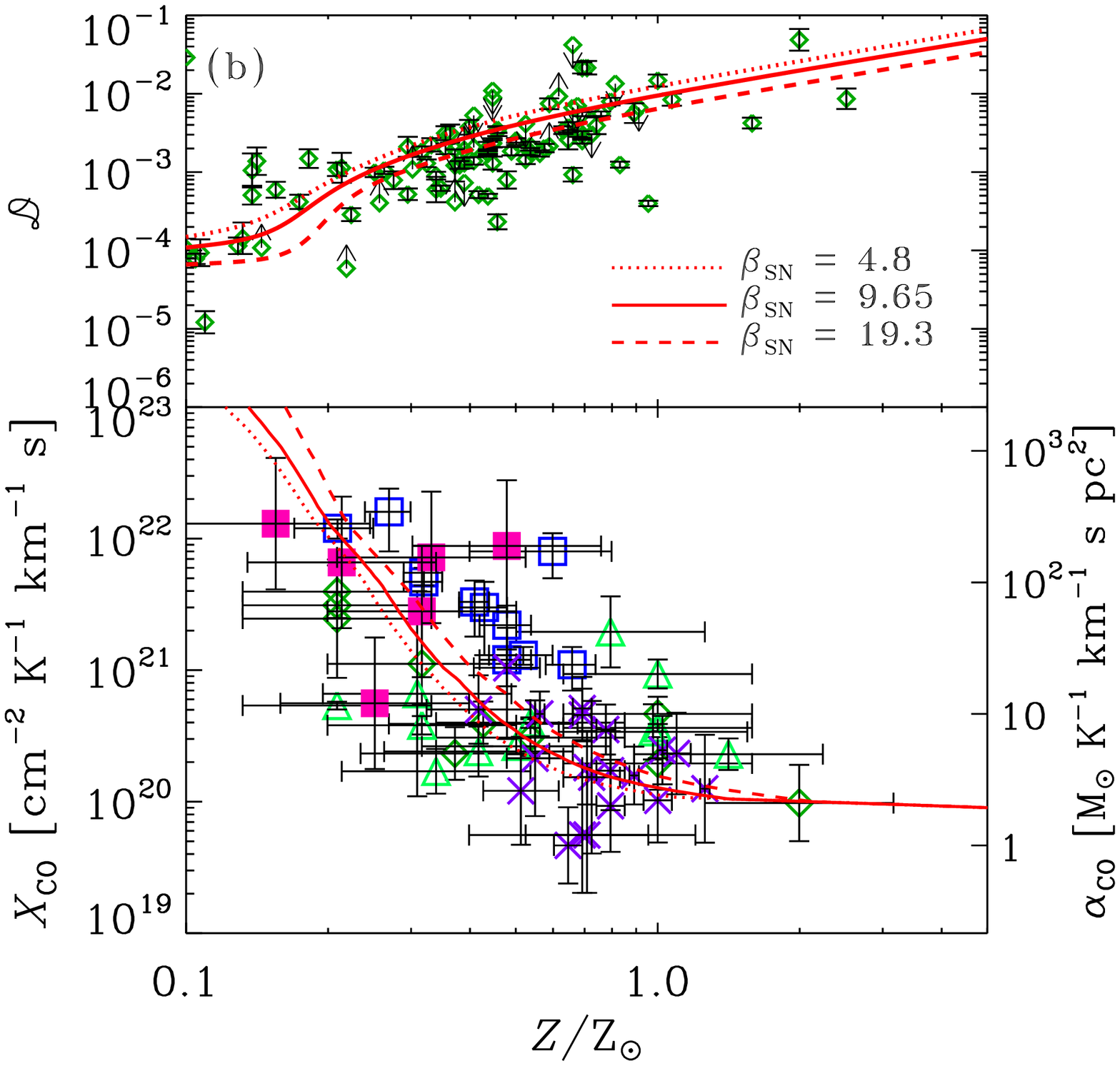}
 \includegraphics[width=0.85\columnwidth]{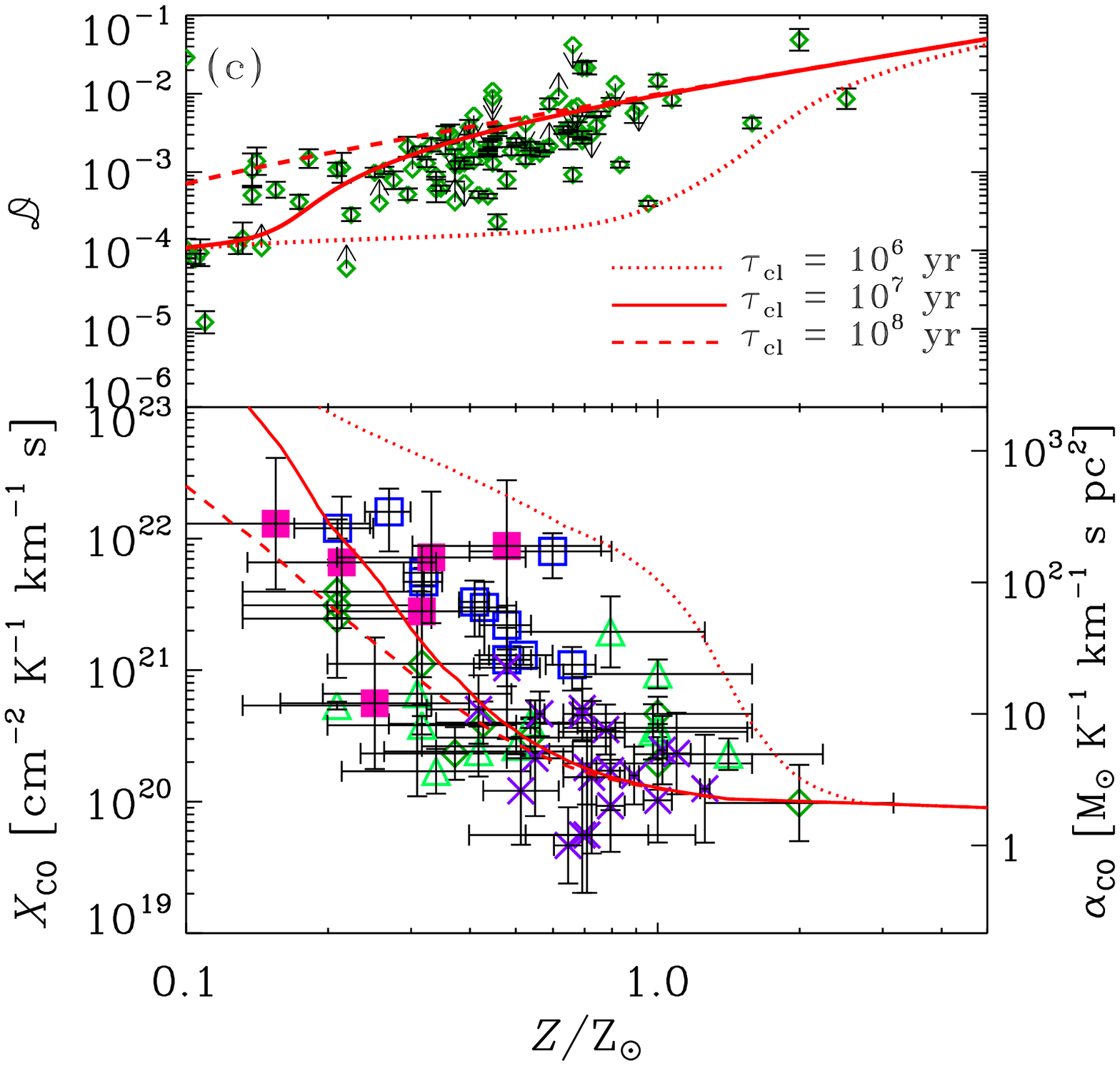}
 \includegraphics[width=0.85\columnwidth]{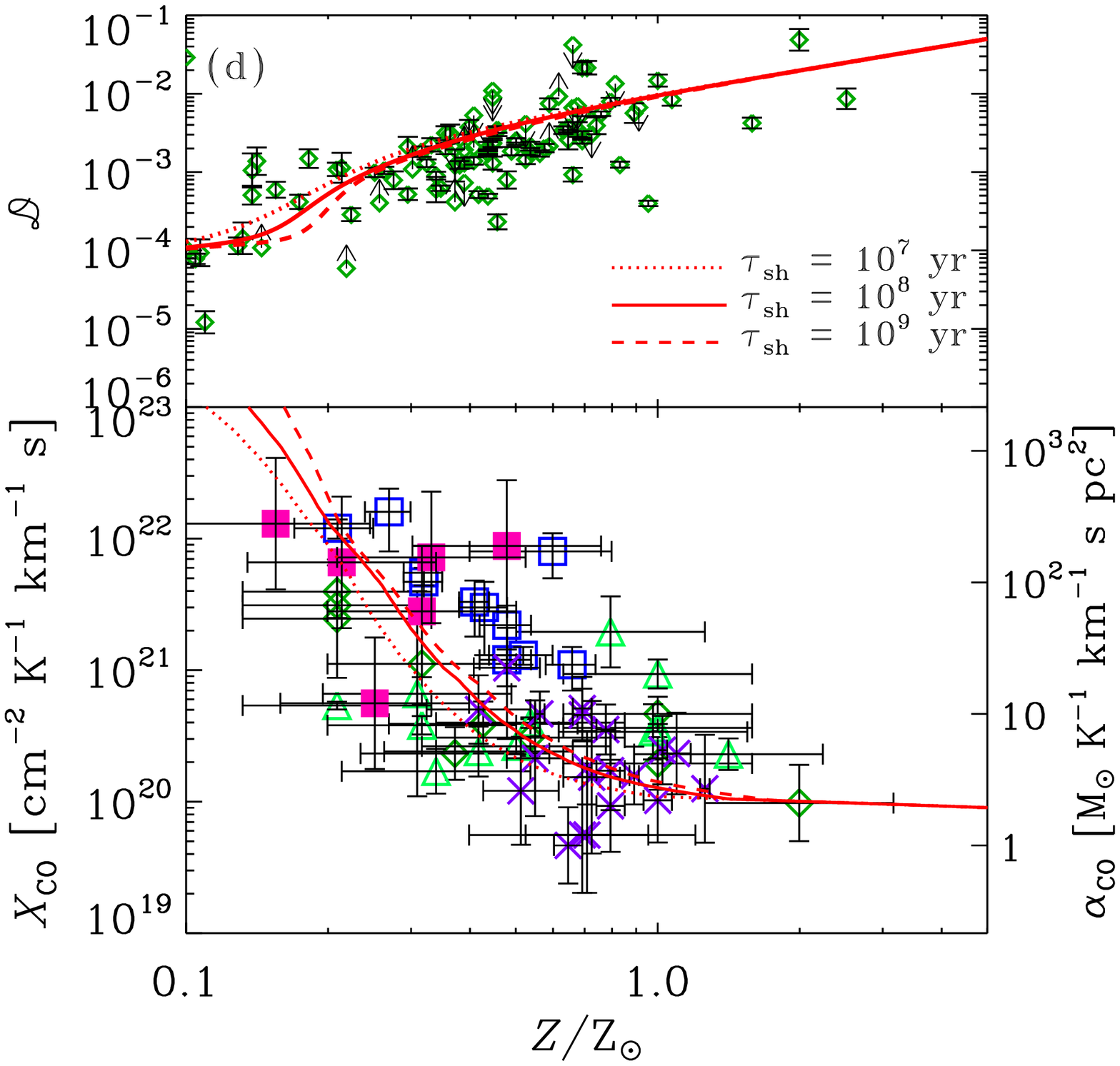}
 \includegraphics[width=0.85\columnwidth]{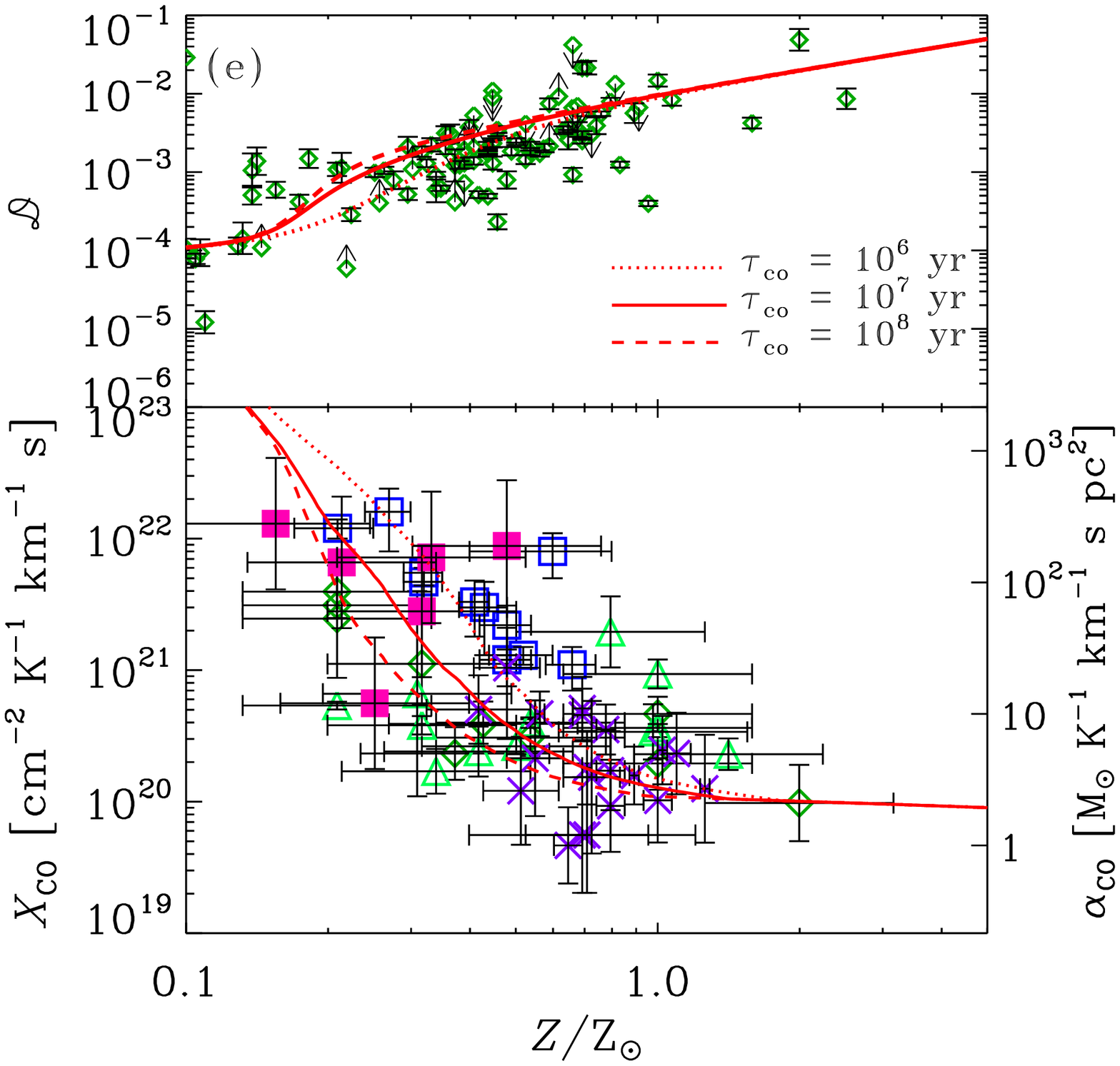}
 \includegraphics[width=0.85\columnwidth]{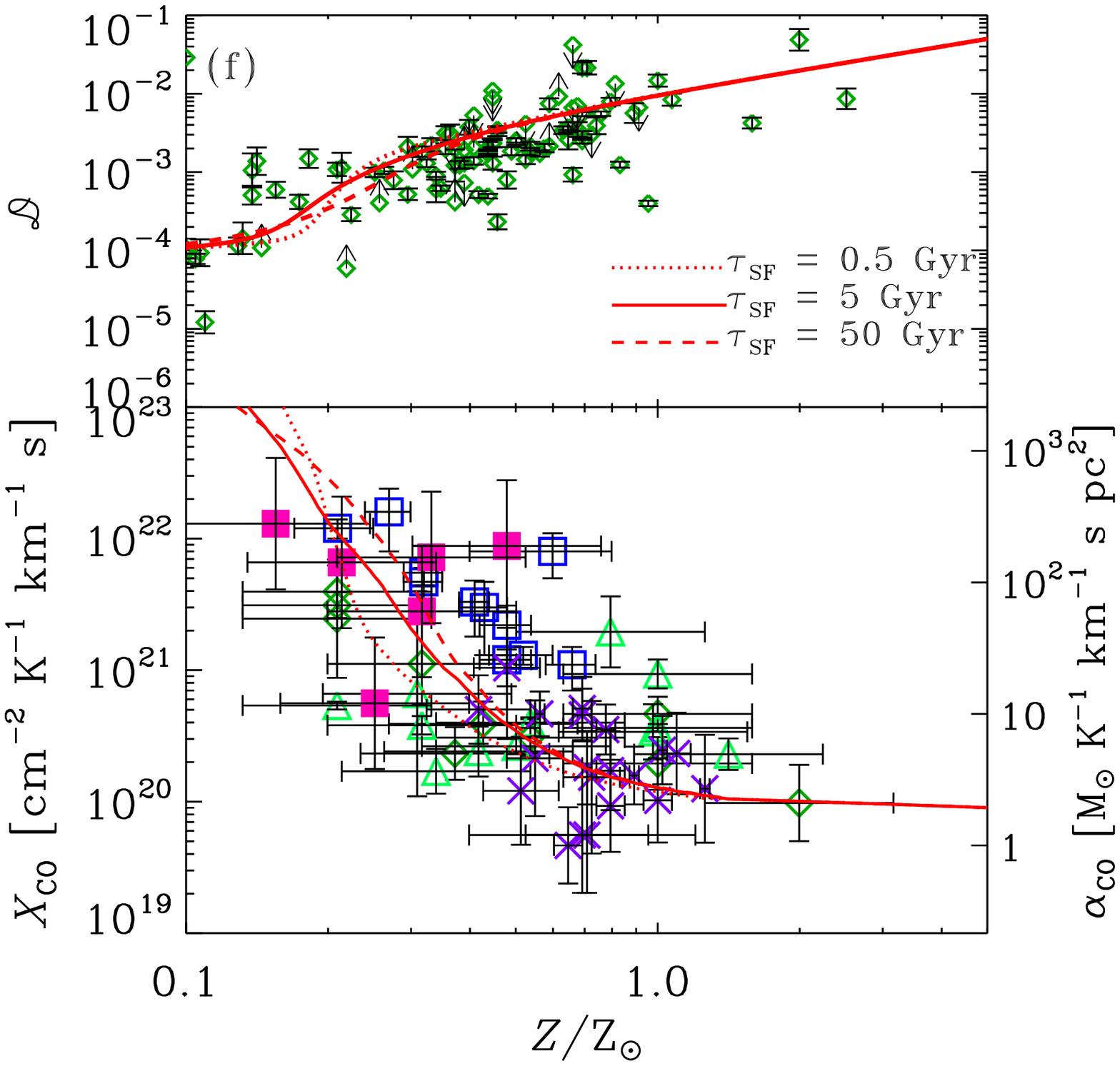}
 \caption{Upper window in each panel: Relation between dust-to--gas ratio and
 metallicity.
 Lower window in each panel: Metallicity dependence of CO-to-H$_2$ conversion factor.
 The two definitions of the conversion factor are shown on the left and
 right vertical axes.
 The models in Panels (a)--(f) are the same as those shown in the top and bottom
 windows of
 Figs.\ \ref{fig:dependence_dust}(a)--(f), respectively. The parameters adopted
 for the lines in each panel are the same as those in
 Fig.\ \ref{fig:dependence_dust}, and are also indicated in
 the panels. We fix the values of
 the parameters other than the varied parameter at the fiducial values listed in Table \ref{tab:param}.
 We adopt the fiducial values for the parameters of the cloud in Table \ref{tab:param_cloud}.
 The points with error bars show the observation data: the data in the upper windows are
 taken from \citet{Remy-Ruyer:2014aa} and those in the lower windows from
 \citet{Leroy:2011aa} (diamonds), \citet{Bolatto:2008aa} (triangles),
 \citet{Israel:1997aa} (open squares), \citet{Sandstrom:2013aa} (crosses), and
 \citet{Cormier:2014aa} (filled squares).
 See Section \ref{subsec:Zdep} for more detailed descriptions of the data.}
 \label{fig:xco}
\end{figure*}

{The other remark is that
the virial-mass-based $X_\mathrm{CO}$ estimate
(i.e.\ \citealt{Bolatto:2008aa} above)
may not be sensitive to CO-dark layers in molecular clouds.
However, we do not find any offset in the $X_\mathrm{CO}$--$Z$
relation between \citet{Bolatto:2008aa}'s sample and the
other samples; thus, we simply include all the above data in the
same diagram, although we should keep in mind the inhomogeneity
in the method of deriving $X_\mathrm{CO}$.
There is another way of observationally deriving $X_\mathrm{CO}$:
\citet{Amorin:2016aa}
derived the molecular gas mass by assuming a star formation law.
The CO-to-H$_2$ conversion factors they derived are consistent with
(or within the scatter of) the ones based on direct gas mass tracers
(i.e.\ the data plotted in Fig.\ \ref{fig:xco}).
We should also note that some galaxies appear for multiple times
and that we did not make any effort of unifying them into a single data
point.
However, the difference among the metallicities of the
same galaxy is within 0.2 dex, which is smaller than the scatter
in the $X_\mathrm{CO}$--$Z$ diagram. Moreover, the
scatter of $X_\mathrm{CO}$ caused by the multiplicity is smaller than
the difference between different galaxies at a similar metallicity;
thus, the multiplicity does not affect our results.}

In Fig.\ \ref{fig:xco}, we show the comparison of our results with
the observational data for the $X_\mathrm{CO}$--$Z$ relation.
We also compare the $\mathcal{D}$--$Z$ relations
with nearby galaxy data taken from \citet{Remy-Ruyer:2014aa}.
{In deriving the dust-to-gas ratio, they estimated the gas mass
by the sum of
atomic and molecular gas mass. For the molecular gas mass,
they examined both the Milky Way  and the metallicity-dependent
CO-to-H$_2$ conversion factors. Strictly speaking we need to
assume a consistent conversion factor with the one predicted
by our model, but they showed that the difference in the
conversion factor only changes the dust-to-gas ratio within
the scatter of the observational data. Thus, we simply adopt
their dust-to-gas ratios based on the Milky Way conversion factor.
The $\mathcal{D}$--$Z$ relations predicted by our models
are broadly consistent with the
data as already discussed in H15.
Our theoretical predictions are also
consistent with the $X_\mathrm{CO}$--$Z$ relation plotted in the figure.}
In general, the scatter of observational
data is too large to constrain most of the parameters in the
models. Nevertheless, inefficient accretion represented by
$\tau_\mathrm{cl}=10^6$ yr can be rejected (see Fig.\ \ref{fig:xco}c),
since it predicts too high conversion
factors. In this case, the dust-to-gas ratio is
so low that shielding of dissociating photons is not enough to realize
a high CO abundance (see also Fig.\ \ref{fig:dependence_dust}c).
Such a low accretion efficiency also tends to underproduce the
observed dust-to-gas ratio as a function of metallicity as
shown in the upper window of Fig.\ \ref{fig:xco}c.

As mentioned above, the scatter
of the observational data is much larger than the variation of
the $X_\mathrm{CO}$--$Z$ relations predicted by the models
if the parameters other than $\tau_\mathrm{cl}$ are changed.
The dust condensation efficiency
in stellar ejecta ($f_\mathrm{in}$) has a large impact at low metallicities, where
CO is difficult to detect. Yet, it is interesting to note that
\citet{Cormier:2014aa}'s lowest metallicity data lie in the
range where the difference in $X_\mathrm{CO}$ among
various values of $f_\mathrm{in}$ can be seen.
At the metallicity range where a statistical number of
data for $X_\mathrm{CO}$ are available,
interstellar processing is more important than stellar dust
production. As concluded by our previous work (H15)
as well as by many other studies
\citep{Dwek:1998aa,Hirashita:1999aa,Inoue:2003aa,Zhukovska:2008aa,Hirashita:2011aa,Valiante:2011aa,Mattsson:2012aa,Kuo:2013aa,de-Bennassuti:2014aa,Michalowski:2015aa,Mancini:2015aa,Schneider:2016aa,Popping:2016aa},
dust growth by accretion is the major driver of the evolution
of dust-to-gas ratio at $Z\ga 0.1$ Z$_{\sun}$
(see \citealt{Rouille:2014aa} for experimental evidence;
but see \citealt{Ferrara:2016aa}).
In conclusion, dust growth by accretion
not only drives the evolution of dust-to-gas ratio but also
have a significant impact on the relation between
$X_\mathrm{CO}$ and metallicity. The other processes
than accretion have minor impacts on the $X_\mathrm{CO}$--$Z$
relation.

\subsection{Effects of grain size distribution}

One of the unique features in our modeling is the capability to
calculate the grain size distribution in the
form of the mass ratio between small and large
grains. The effects of grain size distribution appear
directly in H$_2$ formation (equation \ref{eq:H2_form})
and shielding (dust extinction) (equation \ref{eq:tau_LW}).
We here examine how the grain size distribution affects
these processes.

\subsubsection{Effects on H$_2$ formation}

In order to investigate the effect of grain size distribution on
H$_2$ formation, we take the following two extremes. One is that
we adopt the H$_2$ formation rate on small grains
for both large and small grains in equation (\ref{eq:H2_form});
that is, all the grains are assumed to be small grains when we
estimate the H$_2$ formation rate. The other is the opposite;
that is, we adopt the H$_2$ formation rate of large grains
for both large and small grains. The former extreme represents
the highest H$_2$ formation efficiency (referred to as
the high H$_2$ formation rate), while the latter
shows the lowest efficiency (referred to as the low
H$_2$ formation rate. We also show the H$_2$ formation
rate in our models (consistent with the evolution of
grain size distribution), which is referred to as the standard
H$_2$ formation rate. We use the fiducial values for the
parameters (Tables \ref{tab:param} and \ref{tab:param_cloud})
unless otherwise stated.

\begin{figure*}
 \includegraphics[width=1\columnwidth]{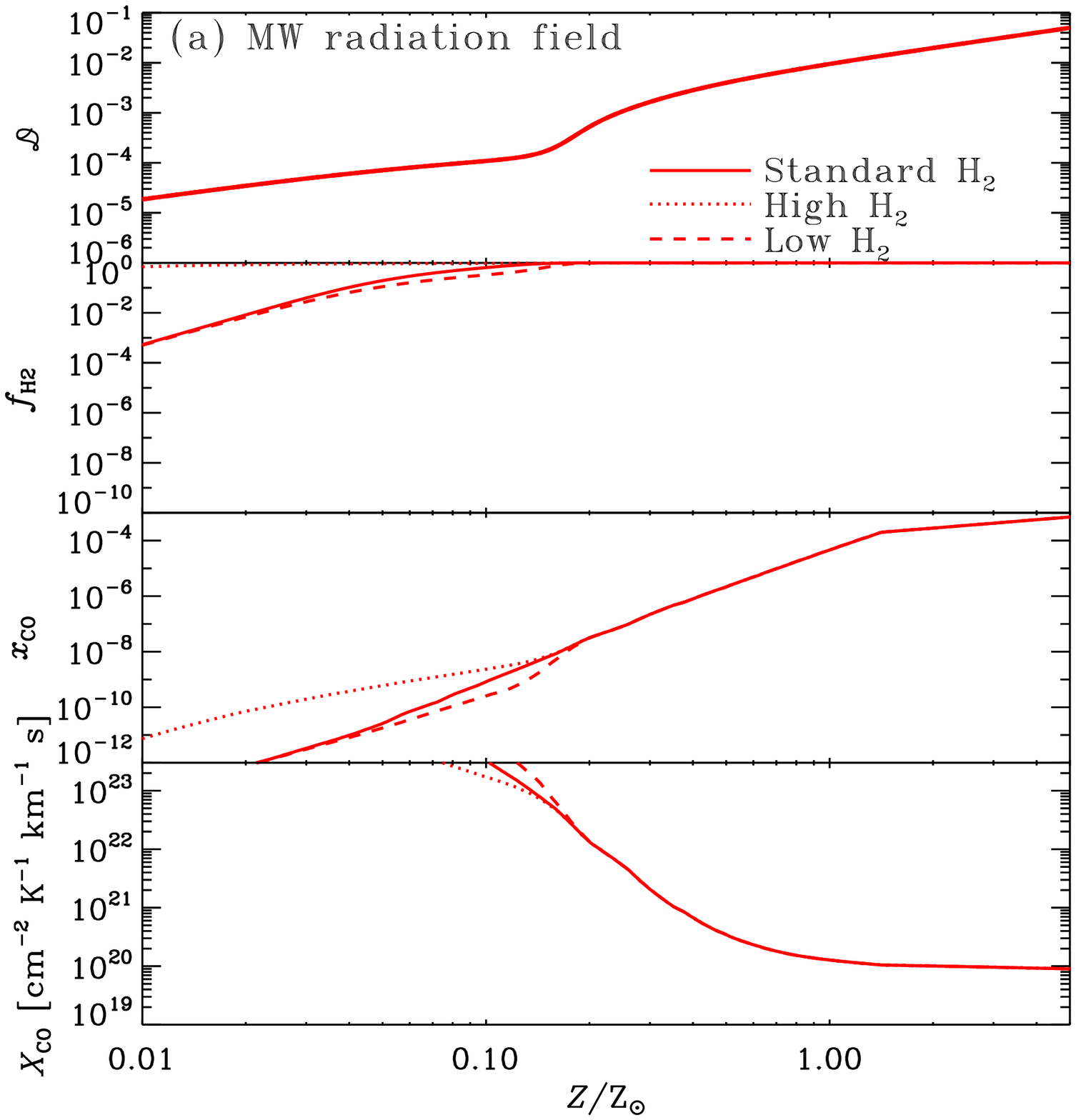}
 \includegraphics[width=1\columnwidth]{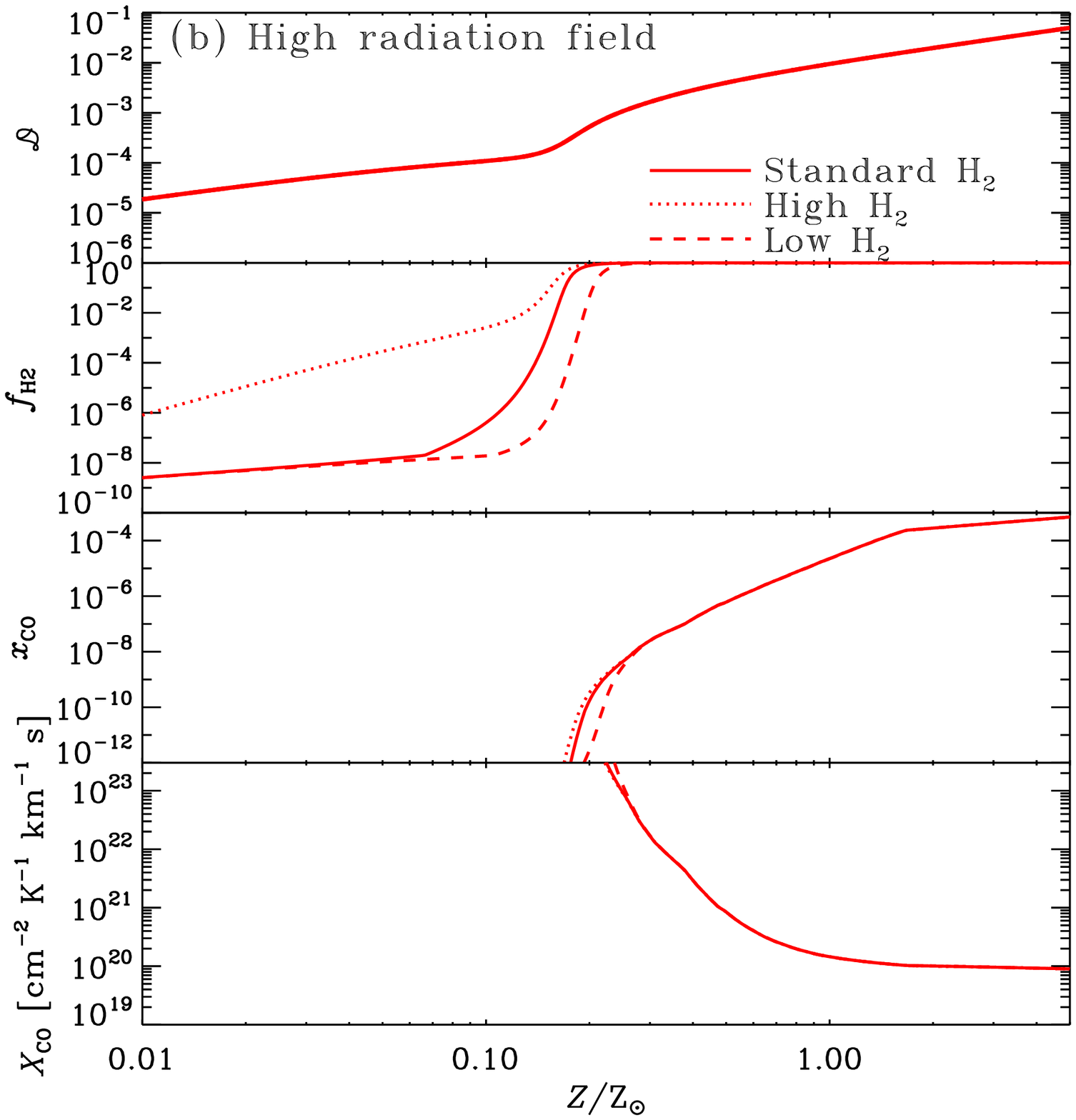}
 \caption{Same as Fig.\ \ref{fig:fiducial} but for high and low H$_2$ formation
 rates, for the dotted and dashed lines, respectively.
 For the high (low) H$_2$ formation rate, we assume that
 all the grains are small (large) grains. Panels (a) and (b) adopt
 $\chi =1.7$ (the Milky Way
 radiation field) and 170, respectively, with the other parameters
 fixed to the fiducial values (Tables \ref{tab:param} and \ref{tab:param_cloud}).}
 \label{fig:H2rate}
\end{figure*}

In Fig.\ \ref{fig:H2rate}a, we show the evolution of the
quantities of interest as a function of metallicity for
the high and low H$_2$ formation rates, in comparison with
the standard H$_2$ formation rate. The high
H$_2$ formation rate predicts that the cloud becomes
almost fully molecular in
the metallicity range $Z\ga 0.01$~Z$_{\sun}$.
The low H$_2$ formation rate shows a similar
evolution to the case of the standard H$_2$ formation rate, since
the dust abundance is dominated by large grains
in the low-metallicity phase of galaxy evolution. The difference
between these two cases become visible around
$Z\sim 0.1$ Z$_{\sun}$ because of the small grain
production by shattering and accretion. Thus, the
enhancement of H$_2$ formation rate by these small
grain production mechanisms is important for the
H$_2$ abundance at such a low metallicity as
$Z\sim 0.1$ Z$_{\sun}$. At higher metallicities,
hydrogen becomes fully molecular, and
the effect of H$_2$ formation rate is not important.
The CO abundance is also affected by different grain
sizes at low metallicities
because of the difference in H$_2$ shielding of
dissociating photons. However, the CO-to-H$_2$
conversion factor does not change significantly by the variation
of grain sizes in the metallicity range
where detection of CO is possible ($Z\ga 0.1$ Z$_\odot$).

As shown in Fig.\ \ref{fig:dependence_nH}, the radiation field
affects the H$_2$ abundance drastically. Thus, we investigate the
effects of grain-size-dependent H$_2$ formation rate for
the strongest radiation field $\chi =170$ in Fig.\ \ref{fig:H2rate}b.
The same behaviour for $f_\mathrm{H_2}$ as seen for
$\chi =1.7$ (the fiducial value) is observed, but with an enhanced
difference among the three cases for the H$_2$ formation rate.
In particular, the H$_2$ fraction is completely different at
$Z\sim 0.1$--0.2 Z$_{\sun}$. However, at higher metallicities,
the gas is fully molecular in any case; thus, the CO abundance
and CO-to-H$_2$ conversion factor are unaffected at
$Z\ga 0.2$ Z$_{\sun}$.

In summary, the effect of grain size distribution on H$_2$ formation
rate is important at low metallicities (typically $Z\la 0.2$~Z$_{\sun}$).
Thus, if we interpret the H$_2$ abundance in low-metallicity systems
such as DLAs \citep{Lanzetta:1989aa,Ledoux:2003aa} and
nearby dwarf galaxies \citep{Kamaya:2001aa}, the effect of grain size
distribution is important, especially for environments with intense
UV radiation field. However, even in the intense UV field,
hydrogen becomes fully molecular at high metallicities
($Z\ga 0.3$ Z$_{\sun}$). Consequently, the CO-to-H$_2$ conversion
factor is not affected by the dependence of H$_2$ formation rate
on the grain size distribution at $Z\ga 0.3$ Z$_{\sun}$.

\subsubsection{Effects on dust shielding}

The grain size distribution also influences dust shielding of
dissociating photons.
As formulated in equations (\ref{eq:shield_dust})
and (\ref{eq:tau_LW}), small grains have larger efficiencies
of shielding than large grains. We take the two extremes:
in one case we adopt the optical depth
for small grains for both large and small grains in
equation (\ref{eq:tau_LW}); that is, all the grains are assumed to be
small when we estimate the optical depth for dissociating photons.
This case is referred to as the high extinction.
The other is the opposite: we adopt the optical depth for
dissociating photons by assuming that all the grains are large grains.
This case is referred to as the low extinction.
The calculation consistent with the grain size evolution is referred to
as the standard extinction. We use the fiducial values for the
parameters (Tables \ref{tab:param} and \ref{tab:param_cloud})
unless otherwise stated.

\begin{figure*}
 \includegraphics[width=1\columnwidth]{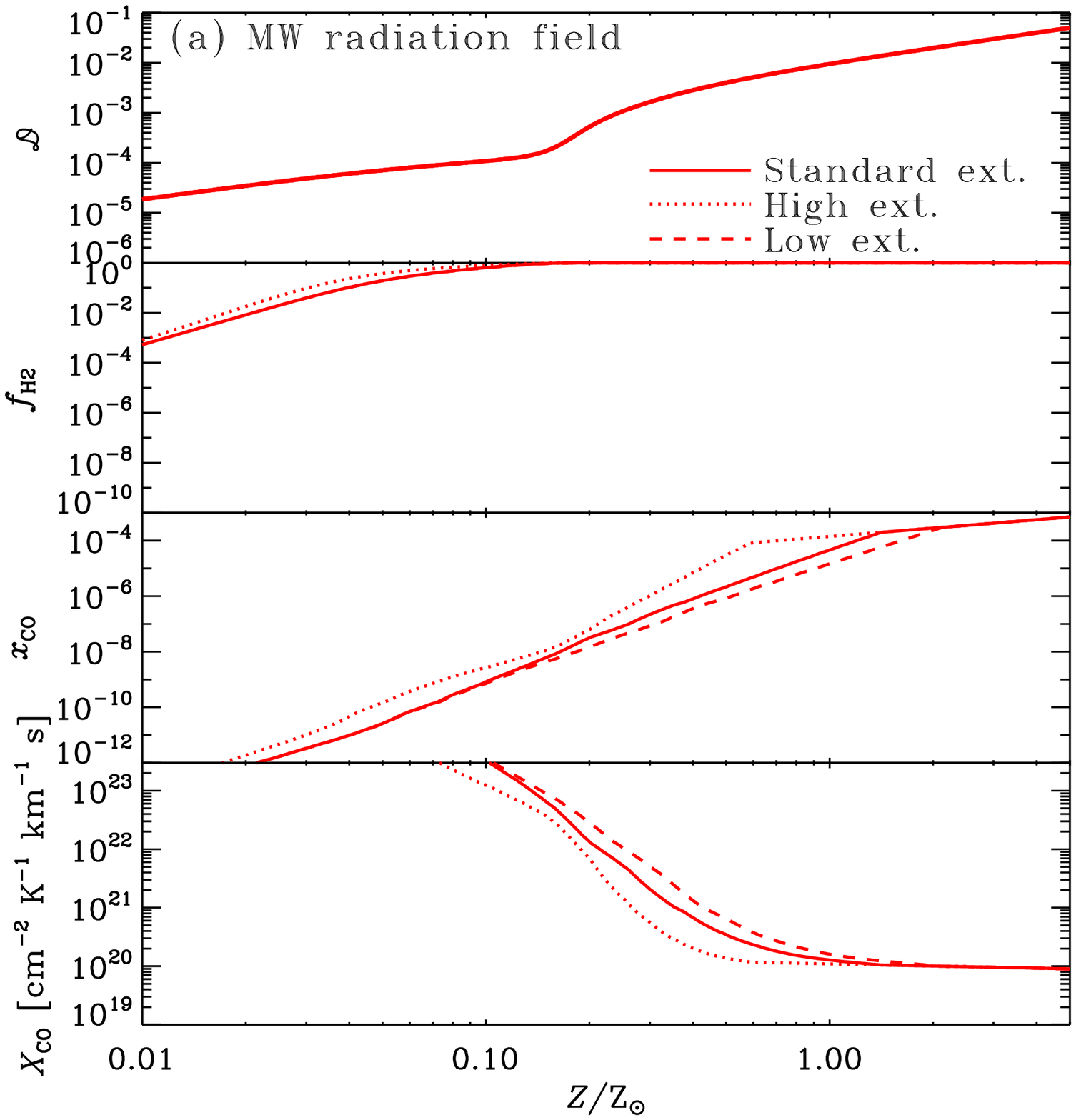}
 \includegraphics[width=1\columnwidth]{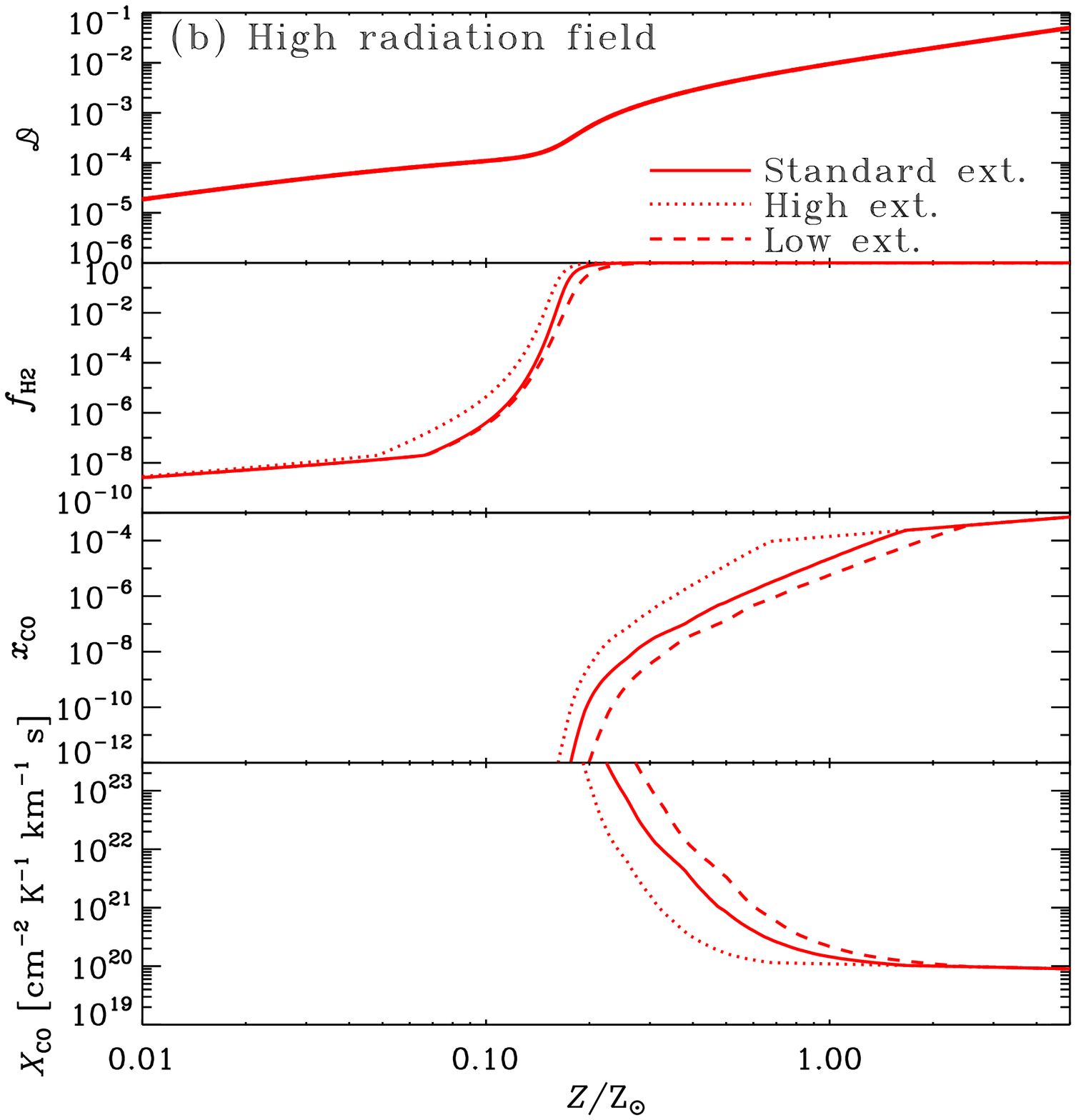}
 \caption{Same as Fig.\ \ref{fig:fiducial} but for high and low dust extinctions
 of dissociating photons, for the dotted and dashed lines, respectively.
 For the high (low) extinctions, we assume that
 all the grains are small (large) grains. Panels (a) and (b) adopt
 $\chi =1.7$ (the Milky Way
 radiation field) and 170, respectively, with the other parameters
 fixed to the fiducial values (Tables \ref{tab:param} and \ref{tab:param_cloud}).}
 \label{fig:ext}
\end{figure*}

In Fig.\ \ref{fig:ext}a, we show the quantities of
interest as a function of metallicity for the high and low extinctions
of dissociating photons in comparison with the standard case.
The change of extinction does not largely affect the H$_2$ fraction
since self-shielding is more important.
The effect of extinction is rather important for the CO fraction and
the CO-to-H$_2$ conversion factor.
In contrast to the effect of grain size on H$_2$ formation,
dust shielding affects the CO abundance
also at high metallicity.

We also investigate a case with high radiation field $\chi =170$
in Fig.~\ref{fig:ext}b. Comparing $f_\mathrm{H_2}$ in
Figs.\ \ref{fig:H2rate}b and \ref{fig:ext}b, we observe that
the effect of grain size on
dust shielding is less important for $f_\mathrm{H_2}$ than
that on H$_2$ formation. On the other hand,
dust shielding is again important for the CO abundance;
in particular, the metallicity dependence of the CO-to-H$_2$
conversion factor at sub-solar metallicity is driven more by dust shielding
than by H$_2$ shielding. This is why the conversion factor is affected by
coagulation and shattering (i.e.\ main mechanisms of changing the grain
size distribution) as shown in Fig.\ \ref{fig:xco}.

\section{Observational implications}\label{sec:implication}

\subsection{Uncertainty in molecular mass estimates}

The variation of CO-to-H$_2$ conversion factor among galaxies
can be produced by the different dust evolution histories.
In particular, we have shown that if the efficiency of
dust growth by accretion varies, the $X_\mathrm{CO}$--$Z$
relation is significantly perturbed.
This causes an uncertainty in molecular mass estimates,
since we do not know the efficiency
of dust growth \textit{a priori} in observing a galaxy.
Note that this uncertainty still remains even if we know the
metallicity.

The molecular gas mass is often related to the star formation rate.
It has been known that there is a good correlation between
these two quantities. The correlation -- the so-called star formation
law -- has been extensively discussed in various contexts
\citep[e.g.][]{Leroy:2013aa}. The above uncertainty caused by the
dust evolution history contributes to the scatter of the observationally
derived H$_2$ mass from CO luminosity. In particular, if a sample
contains objects with a variety of metallicities, the effect of dust evolution,
especially dust growth by accretion, should be carefully discussed,
since it may cause a significant scatter in the $X_\mathrm{CO}$--$Z$
relation as shown in Fig.\ \ref{fig:xco}c.

\subsection{Nearby low-metallicity galaxies}\label{subsec:various_galaxies}

As expected from the above results, low-metallicity galaxies
are indeed deficient in CO \citep[e.g.][]{Schruba:2012aa}, and large part
of carbon atoms are rather traced by [C \textsc{ii}] emission
\citep{Madden:1997aa,Cormier:2014aa}. According to our calculations above,
hydrogen in
a cloud whose column density is typical of molecular clouds
in the local Universe ($N_\mathrm{H}\sim 10^{22}$~cm$^{-2}$)
becomes fully molecular around $Z\sim 0.1$ Z$_{\sun}$,
while the CO-to-H$_2$ conversion factor is still an order of
magnitude larger than the Milky Way value around
$Z\sim 0.4$ Z$_{\sun}$ in our fiducial case. Thus, we expect
that such a fully molecular gas is difficult to be traced by CO around
$\sim$0.1--0.4 Z$_{\sun}$, which just corresponds to
metallicities in nearby low-metallicity dwarf galaxies.
Recently, it has also been suggested
that [C \textsc{i}] is a better tracer of molecular gas at low metallicity
than CO \citep{Glover:2016aa}.
At $Z<0.1$ Z$_{\sun}$, a cloud with
$N_\mathrm{H}\sim 10^{22}$ cm$^{-2}$ is not fully molecular.
Thus, it is expected that H \textsc{i} gas rather than molecular gas
becomes relatively important for the total gas content.

\subsection{Starburst galaxies}\label{subsec:starburst}

\begin{figure}
 \includegraphics[width=1\columnwidth]{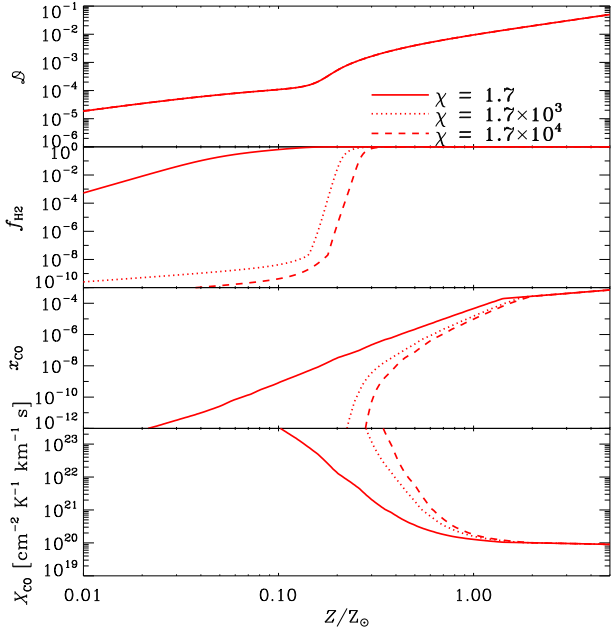}
 \caption{Same as Fig.\ \ref{fig:fiducial} but for extremely high values
 for $\chi$ ($1.7\times 10^3$ and $1.7\times 10^4$; dotted and
 dashed lines) appropriate for
 starburst galaxies. We also show the fiducial value (Milky Way value)
 $\chi =1.7$ (solid line) as a reference. The other parameters are
 fixed to the fiducial values (Tables \ref{tab:param} and \ref{tab:param_cloud}).}
 \label{fig:highchi}
\end{figure}

Some observations have suggested that starburst galaxies
such as (ultra)luminous infrared galaxies
[(U)LIRGs] have
CO-to-H$_2$ conversion factors
$X_\mathrm{CO}\sim 0.2$--$1\times 10^{20}$ cm$^{-2}$ K$^{-1}$ km$^{-1}$ s,
which are lower than the conversion factor in the Milky Way by
a factor of 2--10 \citep[see section 7 of][for a review]{Bolatto:2013aa}.
As argued in Section \ref{subsec:choice},
they have 10--100 times higher surface SFR density than
nearby disc galaxies; thus, we first calculate the metallicity
dependence of the quantities of interest with
$\chi =1.7\times 10^3$ and
$1.7\times 10^4$. The results for these high radiation field intensities
are shown in Fig.\ \ref{fig:highchi}. As expected, the H$_2$ and
CO abundances are suppressed under such a high radiation field.
However, the cloud still becomes fully molecular at
$Z\ga 0.3$ Z$_{\sun}$. The metallicity dependence of
$X_\mathrm{CO}$ is steeper for a higher $\chi$, but
$X_\mathrm{CO}$ is insensitive to $\chi$ around solar
metallicity. This is because dust shielding creates a favourable
condition for the CO formation around solar metallicity even if
the radiation field is four orders of magnitude higher than
the Galactic value. Moreover,
the CO $J=1$--0 line is optically thick at high metallicity
(i.e.\ the metallicity dependence of $X_\mathrm{CO}$ is very weak).
Thus, the observed systematically lower conversion factors in starburst galaxies
should be interpreted not by
metallicity dependence or a stronger radiation field intensity but by a change of other
physical conditions.
In fact, the low conversion factors in starburst galaxies could be
explained by high gas temperatures \citep{Wild:1992aa} or high velocity dispersions
\citep{Zhu:2003aa,Papadopoulos:2012aa}
or both \citep{Narayanan:2011aa}.

\subsection{High-$z$ galaxies}

\citet{Tacconi:2008aa} studied submillimetre galaxies (SMGs)
and UV/optically selected galaxies at $z\sim 2$ and found that
they have similar CO-to-H$_2$ conversion factors to those found
in nearby (U)LIRGs.
\citet{Daddi:2010aa} showed that the so-called main-sequence galaxies
at $z\sim 1.5$
have conversion factors similar to the Milky Way ones.
The difference in the conversion factor between SMGs and
main-sequence galaxies is also found by \citet{Magdis:2011aa}.
\citet{Magnelli:2012aa} show that there is a correlation between
CO-to-H$_2$ conversion factor and dust temperature for galaxies
at $z\sim 1$. Although this dust temperature dependence of $X_\mathrm{CO}$
could be interpreted as an effect of the radiation field,
we have shown above that
the difference in $\chi$, as shown in Fig.\ \ref{fig:dependence_nH},
does not produce a large difference
in the conversion factor around solar metallicity appropriate
for their sample. It should be due to effects not
explicitly included in our paper such as high gas temperatures and large velocity dispersions.

Studies on metallicity dependence of CO-to-H$_2$ conversion factor
have been made possible also at $z\ga 1$.
\citet{Genzel:2012aa} used a sample of main-sequence galaxies
and found a similar metallicity dependence of $X_\mathrm{CO}$
to that seen for nearby galaxies. This indicates that our models,
which explain the $X_\mathrm{CO}$--$Z$ relation in nearby galaxies,
can also be applicable to high-redshift galaxies. Since dust growth by accretion
has a strong influence on the $X_\mathrm{CO}$--$Z$ relation
at a range of $Z$ appropriate for CO detection (i.e.\ $Z\ga 0.1$ Z$_{\sun}$),
their result implies that galaxies at $z\ga 1$ have similar efficiencies of
dust growth by accretion to those in nearby galaxies.

\section{Conclusion}\label{sec:conclusion}

We investigate the effect of dust evolution on the metallicity
dependence of
CO-to-H$_2$ conversion factor ($X_\mathrm{CO}$).
To this aim, we first need to understand the dependence
of dust abundance on metallicity, since dust affects the molecular
abundances through H$_2$ formation on dust surface and
shielding of dissociating photons.
The dust abundance as a function of metallicity is modeled
by including all major processes driving the dust evolution:
dust condensation in stellar ejecta, dust destruction in
SN shocks, grain growth by accretion and coagulation,
and grain disruption by shattering. The grain size distribution is
represented by the small-to-large grain abundance ratio, which is
consistently treated with the dust evolution.
The relation between
dust-to-gas ratio and metallicity ($\mathcal{D}$--$Z$ relation)
is used to calculate the H$_2$ fraction and CO abundance
in a cloud with $N_\mathrm{H}\sim 10^{22}$ cm$^{-2}$
as a function of metallicity by considering  dust shielding of
dissociating photons and H$_2$ formation on dust surface
under given interstellar radiation field and hydrogen column density.
The effects of grain size on
H$_2$ formation and shielding are also included.
The major output of our model is
the CO-to-H$_2$ conversion factor, $X_\mathrm{CO}$.

As a consequence of the modeling,
{we predict consistent metallicity dependence
of $X_\mathrm{CO}$ ($X_\mathrm{CO}$--$Z$ relation) with
observational data}. Among various processes driving dust evolution,
grain growth by accretion has the largest
impact on the $X_\mathrm{CO}$--$Z$ relation.
Efficient accretion, which is also needed to reproduced the
$\mathcal{D}$--$Z$ relation of nearby galaxies, is strongly required
to explain the $X_\mathrm{CO}$--$Z$ relation.
The dust condensation efficiency in stellar ejecta also affects
$X_\mathrm{CO}$, but the effect appears at low metallicity
($\la 0.2$~Z$_{\sun}$), where detection of CO is difficult.
The other processes also have some impacts on $X_\mathrm{CO}$,
but the effects are minor compared with the scatter of the observational data
at the metallicity range ($Z\ga 0.1$ Z$_{\sun}$) where CO could be detected.

We also find that dust condensation in stellar ejecta has a dramatic
impact on the H$_2$ abundance at low metallicities
($\la 0.1$~Z$_{\sun}$) and that the grain size dependence of H$_2$ formation
rate is also important. Such a metallicity range is relevant for
damped Lyman $\alpha$ systems and extremely low-metallicity dwarf
galaxies. Between $\sim 0.1$ and $\sim 0.4$ Z$_{\sun}$, although
clouds with a typical column density of `molecular clouds'
($N_\mathrm{H}\sim 10^{22}$ cm$^{-2}$) become fully molecular ($f_\mathrm{H_2}\sim 1$),
$X_\mathrm{CO}$ is more than two orders of magnitude larger
than the Milky Way value; thus, molecular clouds
are CO-dark in this metallicity range. Applicability of our
models to the $X_\mathrm{CO}$--$Z$ relation of
$z\ga 1$ main-sequence galaxies implies that their dominant
mechanism of dust abundance evolution is similar to that in nearby
galaxies.

\section*{Acknowledgements}

We are grateful to A. R\'{e}my-Ruyer for providing us with
the data for the relation between dust-to-gas ratio and metallicity
of nearby galaxies. HH thanks A. Ferrara, D. Cormier, R. Feldmann,
and A. D. Bolatto for extremely useful discussions and comments
and the staff at Kavli Institute for Theoretical Physics, University of
California, Santa Barbara for their hospitality during the program
`the Cold Universe'.
We also thank the anonymous referee for useful comments.
This research was supported in part by the National Science
Foundation under Grant No.\ NSF PHY11-25915.
HH is supported by the Ministry of Science and Technology
grant MOST 105-2112-M-001-027-MY3.



\bibliographystyle{mnras}
\bibliography{hirashita}


\appendix

\section{Assumed grain size distribution and moments}
\label{app:moment}

In the two-size approximation, we simply represent the
grain sizes with large and small grains. However, in order to
predict statistical properties, it is
necessary to \textit{assume} a functional form for the grain size distribution.
Although adopting a specific functional form for the grain size distribution
is not essential as mentioned below, we explain how to calculate
statistical properties.

For convenience, we introduce the $\ell$th moment of grain size as
\begin{align}
\langle a^\ell\rangle_i=\frac{1}{n_{\mathrm{d},i}}\int_0^\infty
a^\ell n_i(a)\,\mathrm{d}a,
\end{align}
where $n_{\mathrm{d},i}$ is the number density of dust component $i$
(small or large grains) estimated as
\begin{align}
n_{\mathrm{d},i}=\int_0^\infty n_i (a)\,\mathrm{d}a.
\end{align}
The normalization of grain size distribution is written as
\begin{align}
\frac{4}{3}\pi\langle a^3\rangle sn_{\mathrm{d},i}=
\mu m_\mathrm{H}n_\mathrm{H}\mathcal{D}_i,
\end{align}
where $s$ is the material density of dust,
$n_\mathrm{H}$ is the number density of
hydrogen nuclei, and $m_\mathrm{H}$ is the atomic
mass of hydrogen. For this normalization, we adopt
$s=3.3$ g cm$^{-3}$ \citep{Draine:1984aa} and $\mu =1.4$.
We also need to fix a functional form for $n_i(a)$: we adopt the
modified-lognormal form proposed in H15; that is,
\begin{align}
n_i(a)=\frac{C_i}{a^4}\exp\left\{ -\frac{[\ln (a/a_{0,i})]^2}{2\sigma^2}\right\} ,
\end{align}
where we adopt $a_\mathrm{0,s}=0.005~\micron$,
$a_\mathrm{0,l}=0.1~\micron$ for the central grain radius for
small and large grains, respectively, and $\sigma =0.75$.
The assumption on
grain size distribution is only for the purpose of calculating
quantities that reflects the statistical properties of grain
size distribution such as the above moments and extinction curves.
We calibrate $a_{0,i}$ and $\sigma$ so that the Milky Way extinction
curve is reproduced with $\mathcal{D}_\mathrm{l}=\mathcal{D}_\mathrm{MW,l}=0.007$
and $\mathcal{D}_\mathrm{s}=\mathcal{D}_\mathrm{MW,s}=0.003$.
The assumption on the functional form is not essential after such a
calibration. For convenience, we list the moments in Table \ref{tab:moments}.
Since the volume-to-surface ratio is important for H$_2$ formation,
we also list $\langle a^3\rangle /\langle a^2\rangle$.

\begin{table}
\centering
\begin{minipage}{80mm}
\caption{Moments.}
\label{tab:moments}
\begin{center}
\begin{tabular}{@{}lcccc} \hline
 & $\langle a\rangle$ & $\langle a^2\rangle$ & $\langle a^3\rangle$
 & $\langle a^3\rangle /\langle a^2\rangle$\\
 & [$\micron$] & [$\micron^2$] & [$\micron^3$] & [$\micron$]
\\ \hline
Small grains & $1.2\times 10^{-3}$ & $2.6\times 10^{-6}$ & $9.9\times 10^{-9}$
& $3.8\times 10^{-3}$\\
Large grains & $2.5\times 10^{-2}$ & $1.1\times 10^{-3}$ & $8.0\times 10^{-5}$
& $7.6\times 10^{-2}$\\ \hline
\end{tabular}
\end{center}
\end{minipage}
\end{table}

\section{Calculation of Accretion Efficiency $\beta_\mathrm{acc}$}
\label{app:beta_acc}

We summarize the formula for the efficiency of
grain growth by accretion ($\beta_\mathrm{acc}$) in
H15 \citep[see also][]{Hirashita:2011aa}.
We refer the interested reader to H15 for the detailed derivation.
The accretion time-scale ($\tau_\mathrm{acc}$) is related to the lifetime of cold cloud
(cold clouds mean the gas hosting grain growth by accretion) as
$\tau_\mathrm{acc}=\tau_\mathrm{cl}/(\mathcal{B}X_\mathrm{cl}$)
where $X_\mathrm{cl}$ is the cold cloud fraction to the total
gas mass, $\tau_\mathrm{cl}$ is the lifetime of the cold clouds, and
$\mathcal{B}$ is the increment of dust mass in the cold
clouds, which can
be estimated as
\begin{align}
\mathcal{B}\simeq\left[
\frac{\langle a^3\rangle_\mathrm{s}}{3y\langle a^2\rangle_\mathrm{s}
+3y^2\langle a\rangle_\mathrm{s}+y^3}
+\frac{\mathcal{D}_\mathrm{s}}{Z-\mathcal{D}}\right]^{-1},\label{eq:beta}
\end{align}
where
$y\equiv a_0\xi\tau_\mathrm{cl}/\tau$ [$a_0$ is just used
for normalization,
$\xi\equiv (Z-\mathcal{D})/Z$ is the fraction of metals in the gas phase,
and $\tau$ is the accretion time-scale for a grain with
radius $a_0$], and
$\langle a^\ell\rangle_\mathrm{s}$ is the $\ell$th moment of grain radius
for small grains.
We adopt the following expression for $\tau$:
\begin{align}
\tau=6.3\times 10^7\left(\frac{Z}{\mathrm{Z}_{\sun}}\right)^{-1}
a_{0.1}n_3^{-1}T_{50}^{-1/2}S_{0.3}^{-1}~\mathrm{yr},\label{eq:tau}
\end{align}
where $a_{0.1}\equiv a_0/(0.1~\micron )$,
$n_3\equiv n_\mathrm{H}/(10^3~\mathrm{cm}^{-3})$
($n_\mathrm{H}$ is the number density of hydrogen nuclei in
the cold clouds; we adopt $n_\mathrm{H} = 10^3$~cm$^{-3}$),
$T_\mathrm{50}\equiv T_\mathrm{gas}/(50~\mathrm{K})$ ($T_\mathrm{gas}$
is the gas temperature; we adopt $T_\mathrm{gas} = 50$ K), and
$S_{0.3}\equiv S/0.3$ ($S$ is the
sticking probability of the dust-composing material onto the preexisting
grains; we adopt $S=0.3$).
Assuming that the cold clouds hosting accretion and star formation
are the same, we can express the star formation time-scale as
$\tau_\mathrm{SF}=\tau_\mathrm{cl}/(\epsilon X_\mathrm{cl})$,
where $\epsilon$ is the star formation efficiency of the
cold clouds. We define
$\beta_\mathrm{acc}\equiv\tau_\mathrm{SF}/\tau_\mathrm{acc}$,
which can be evaluated as
\begin{align}
\beta_\mathrm{acc}=\frac{\mathcal{B}}{\epsilon}.
\end{align}
We assume $\epsilon =0.1$ in this paper.

\bsp	
\label{lastpage}
\end{document}